\documentclass{ws-spin}

\usepackage{graphicx} 
\usepackage{multicol}
\usepackage{cite}
\usepackage{float}
\usepackage{color}
\usepackage{dcolumn}  

\begin{document}

\markboth{G. H. Fecher, et al.}{Spectroscopy of Co$_2$MnSi thin films}

\title{State of Co and Mn in half-metallic ferromagnet Co$_2$MnSi 
       explored by magnetic circular dichroism in hard X-ray photoelectron
       emission and soft X-ray absorption spectroscopies}
      
\author{Gerhard~H.~Fecher}
\address{Max Planck Institute for Chemical Physics of Solids, \\
         01187 Dresden, Germany. \\
\email{fecher@cpfs.mpg.de} }

\author{Daniel~Ebke, Siham~Ouardi, Stefano~Agrestini, Chang-Yang~Kuo, 
        Nils~Hollmann, Zhiwei~Hu}
\address{Max Planck Institute for Chemical Physics of Solids, \\
         01187 Dresden, Germany. } 
              
\author{Andrei~Gloskovskii}
\address{Deutsches Elektronen-Synchrotron DESY, \\ 
         22607 Hamburg, Germany}
              
\author{Flora~Yakhou, Nicholas~B.~Brookes}
\address{European Synchrotron Radiation Facility (ESRF), \\ 
         CS40220, F-38043 Grenoble Cedex 9, France}
        
\author{Claudia~Felser}
\address{Max Planck Institute for Chemical Physics of Solids, \\
         01187 Dresden, Germany. }        
        
\maketitle

\begin{history}
\received{Day Month Year}
\revised{Day Month Year}
\end{history}

\begin{abstract}

The half-metallic Heusler compound Co$_2$MnSi is a very attractive material for 
spintronic devices because it exhibits very high tunnelling magnetoresistance 
ratios. This work reports on a spectroscopic investigation of thin Co$_2$MnSi 
films as they are used as electrodes in magnetic tunnel junctions. The 
investigated films exhibit a remanent in-plane magnetisation with a magnetic 
moment of about 5~$\mu_B$ when saturated, as expected. The low coercive field of 
only 4~mT indicates soft magnetic behaviour. Magnetic dichroism in emission and 
absorption was measured at the Co and Mn $2p$ core levels. The photoelectron 
spectra were excited by circularly polarised hard X-rays with an energy of of 
6~keV and taken from the remanently magnetised film. The soft X-ray absorption 
spectra were taken in an induction field of 4~T. Both methods yielded large 
dichroism effects. An analysis reveals the localised character of the electrons 
and magnetic moments attributed to the Mn atoms, whereas the electrons related 
to the Co atoms contribute an itinerant part to the total magnetic moment.

\end{abstract}


\begin{multicols}{2}
\newpage
\section{Introduction} 

Half-metallic ferromagnets (HMFs) have attracted increasing interest because of 
their possible applications in spintronics devices. Their use has been suggested 
to realise spin-dependent devices such as spin injectors, tunnelling 
magnetoresistance (TMR) junctions, and giant magnetoresistance devices. In HMF 
materials, the minority spin density of states (DOS) exhibits a band gap at the 
Fermi energy $\epsilon_F$, while the majority spins are responsible for the 
metallic properties. Among them, Co$_2$-based Heusler alloys with $L2_1$ 
structure [prototype Cu$_2$MnAl, cF16, space group $F\:m\overline{3}m$ (225)] 
have attracted attention for future application as ferromagnetic electrodes in 
spintronic devices\cite{KTH04}. In particular, Co$_2$MnSi, with a high Curie 
temperature of 985~K and large magnetic moment of 5~$\mu_B$, has been explored 
over a long period~\cite{web71,RRW01,GBW02,SBK06}. Very recently, the half-metallicity 
of Co$_2$MnSi thin films was directly reported. A spin polarisation 
of 93\% at room temperature was measured by {\it in-situ} spin-resolved 
ultraviolet photoelectron spectroscopy~\cite{JMB14}. Several groups developed 
fully epitaxial magnetic tunnel junctions (MTJs) based on Co$_2$MnSi as a lower 
electrode and an AlO$_x$~\cite{SNO05a,SHO06} or MgO~\cite{IMK06,IHM08} 
tunnelling barrier. The TMR ratio of Co$_2$MnSi-based MTJs  was improved from 
570\%~\cite{SHO06} to 1135\%~\cite{YIT10} at 4.2~K. Systematic studies of the 
effect of nonstoichiometry on the spin-dependent tunnelling characteristics of 
Co$_2$Mn$_\alpha$Si-based MTJs have shown a direct dependence of the TMR ratio 
on the Mn content~$\alpha$. Recently, giant TMR ratios of up to 1995\% at 4.2~K 
and up to 354\% at 290~K were obtained for epitaxial off-stoichiometric 
Co$_2$Mn$_{1.35}$Si$_{0.88}$/MgO/Co$_2$Mn$_{1.35}$Si$_{0.88}$ MTJs~\cite{LHT12}. 
It is incontrovertible that the Mn content influences the TMR ratio in 
Co$_2$MnSi-based MTJs. To understand the Mn behaviour in this type of Heusler 
compound, more element-specific investigation is required. 

Since the observation of X-ray magnetic circular dichroism (XMCD) from Fe with 
hard X-rays ($K$ edge) by Sch{\"u}tz {\it et al.}~\cite{SWW87} and of Ni with 
soft X-rays ($L$ edges) by Chen~\cite{CSM90}, XMCD measurements are widely used 
as a variant of X-ray absorption spectroscopy (XAS) to investigate the magnetic 
properties of complex systems in an element-specific way. XMCD enables the 
determination of the magnetic properties of the constituent elements of 
materials, as well as their spin and orbital magnetic moments, using integral 
sum rules~\cite{vdL91}. XMCD investigation of several Heusler compounds and thin 
films verified the localised character of the Mn $3d$ valence 
states~\cite{SMH05,TKL06}. The $L_{3,2}$ edge absorption spectra for left and 
right circularly polarised soft X-rays reflect the spin-resolved partial density 
of states (PDOS) at the $3d$ transition metal atoms. XMCD studies of epitaxial 
Heusler thin films of Co$_2$Mn$_{1-x}$Fe$_x$Si provided a pathway for the 
improvement of HMF materials and interfaces for spintronic devices using a 
calculation scheme for recovering the spin-resolved unoccupied Co 
DOS~\cite{KKS09}. In contrast to photon absorption, photoelectron spectroscopy 
allows a detailed illustration of the occupied states (valence band) close to 
the Fermi energy.

Hard X-ray photoelectron spectroscopy (HAXPES) has evolved into the most 
relevant, powerful, and nondestructive method of investigating the bulk 
electronic structure of solids, thin films, and multilayers. Owing to the large 
electron escape depth (the electron mean free path at kinetic energies of 6 keV 
is about 17~nm), it was possible to detect the valence band of buried Co$_2$MnSi 
Heusler compounds underneath 20~nm MgO, which was comparable with those of the 
bulk material~\cite{FBG08,OFF13}. The combination of HAXPES with variable photon 
polarisation provides, in addition to the electronic structure, the magnetic 
structures of buried layers. Very recently, magnetic circular dichroism in the 
angular distribution of photoelectrons (MCDAD) from core states provided another 
hint regarding the localised character of the magnetic moments of some 
transition elements, such as Fe and Co~\cite{KFS11}. In FeGd, MCDAD of the rare 
earth element Gd reflects a stronger temperature dependence of the Gd moment 
compared to that of Fe~\cite{ECM13}.

The present study reports on a detailed investigation of the magnetic structure 
of epitaxial Co$_2$MnSi thin films. The spin-resolved DOSs are calculated using 
full potential {\it ab-initio} methods. Circularly polarised radiation in 
combination with bulk-sensitive HAXPES is used to study the magnetic circular 
dichroism of the core states. The element-specific magnetic moments are 
investigated by circular dichroism in XAS.

\section{Experimental details} 
\label{sec:exp}

Half of a stack for MTJs was prepared by DC/RF sputtering with 4" magnetrons. 
The layers were deposited on MgO substrates in the following order: \\
MgO(001)/MgO\{5\}/Co$_2$MnSi\{20\}/MgO\{2\}.  \\
The numbers in braces give the thickness of the layers in nm. A 5~nm thick MgO layer was 
deposited onto the MgO(001) substrate to level out defects of its surface. This 
results in an improved structure of the Heusler layer. All the layers of the 
films were deposited at room temperature. The argon process pressure was set to 
$1.5\times 10^{-3}$~mbar in the sputtering system, where the base pressure was 
about $10^{-7}$~mbar. The samples were post-annealed for 1~h at $400^{\circ}$C 
in vacuum ($2\times 10^{-7}$~mbar).

The magnetic properties of the films were determined at room temperature using an 
alternating-gradient magnetometer. The films exhibit an in-plane magnetic 
anisotropy. The measured hysteresis loop is shown in Figure~\ref{fig:magmom}. The 
saturation moment is 5~$\mu_B$, as expected. The films are rather soft magnetic 
with a coercive field of only 4~mT. The magnetic moment in remanence is 
2.4~$\mu_B$, which is about half of the saturation moment.

\begin{figurehere}
   \centerline{\psfig{file=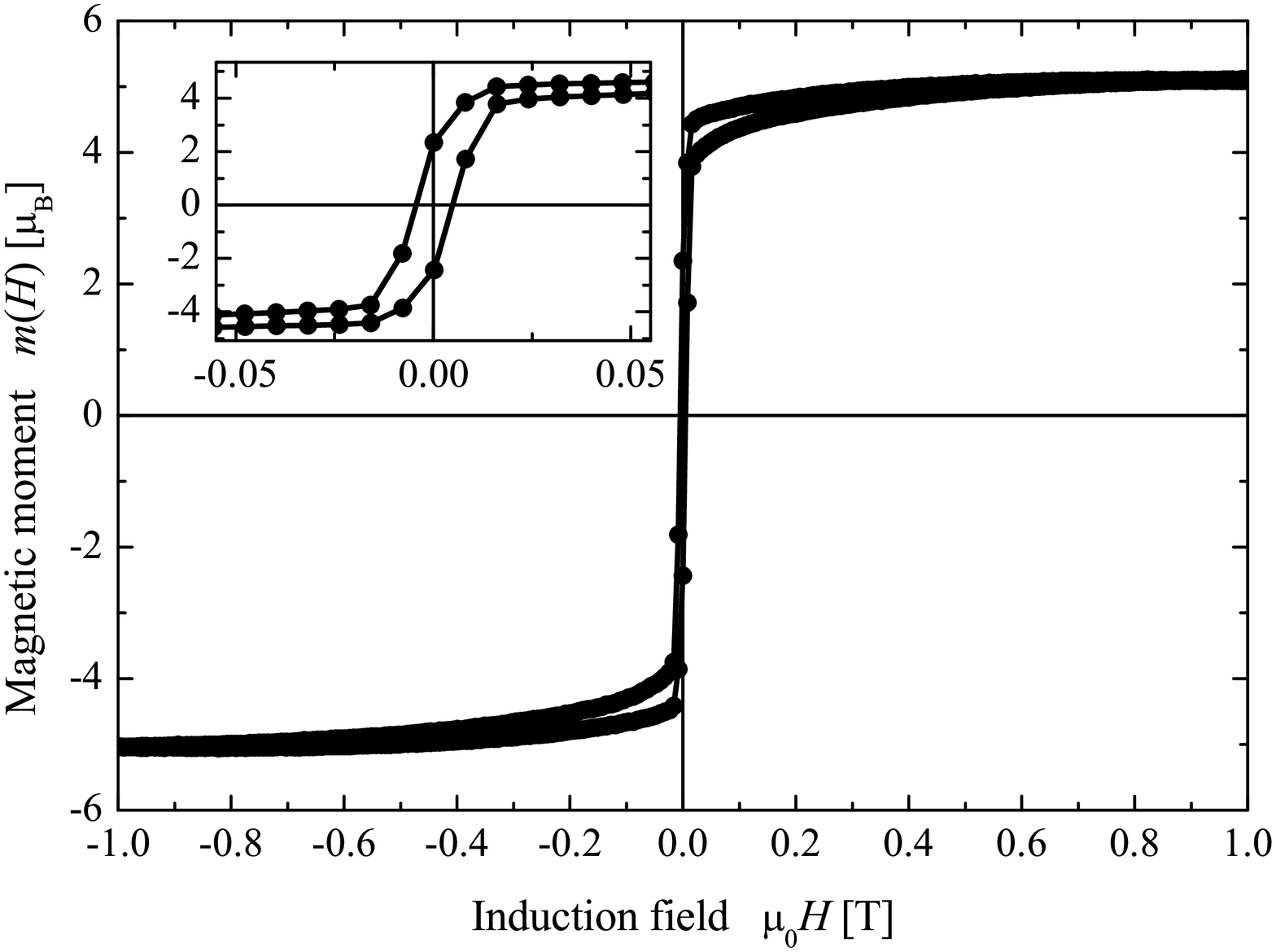,width=8cm}}
   \caption{Magnetisation of Co$_2$MnSi thin films.\\
            The hysteresis was measured at room temperature
            using an alternating-gradient magnetometer. }
\label{fig:magmom}
\end{figurehere}

The HAXPES measurements were made at beamline P09~\cite{SFR13} of PETRA~III 
(Hamburg). See Reference~\cite{GSF12} for the details of the HAXPES setup. The 
photon energy was set to $h\nu = 5947.9$~eV using a Si(111) high heat load 
double-crystal primary monochromator and the (333) reflection of a Si double 
channel-cut post monochromator. The resolution of the monochromator is better 
than 100~meV. The magnetic circular dichroism was measured at a fixed 
magnetisation by changing the helicity of the photons. The thin films were 
magnetised {\it in situ} along the direction of the photon beam after they were 
introduced into the ultrahigh-vacuum (UHV) $\mu$-metal chamber. The helicity of 
the photons was changed using the (111) reflection of an in-vacuum phase 
retarder based on a 400~$\mu$m thick diamond single crystal with (100) 
orientation~\cite{Fra13}. The degree of circular polarisation is about 98 to 
99\%. The energy in the spectra is given with respect to the Fermi energy 
$\epsilon_F$, which was calibrated from measurements of a Au sample. 
$\epsilon_F$ appeared at a kinetic energy of $E_{kin}=5946.48$~eV with a width 
of 200~meV. Accounting for the 28~meV thermal broadening of the Fermi--Dirac 
distribution at room temperature (300~K), this corresponds to an overall energy 
resolution of about 170~meV ($E/\Delta E \approx 3.5\times 10^4$). The setup of 
the HAXPES-MCDAD experiment is sketched in Figure~\ref{fig:setup_mcdad}. A 
retractable Fe-Nd-B permanent magnet was used to magnetise the sample remanently 
or change the direction of magnetisation. It supplies an induction field of 
about $\pm1$~T, which is high enough to saturate the magnetisation of the thin 
film (compare Figure~\ref{fig:magmom}).

\begin{figurehere}
   \centerline{\psfig{file=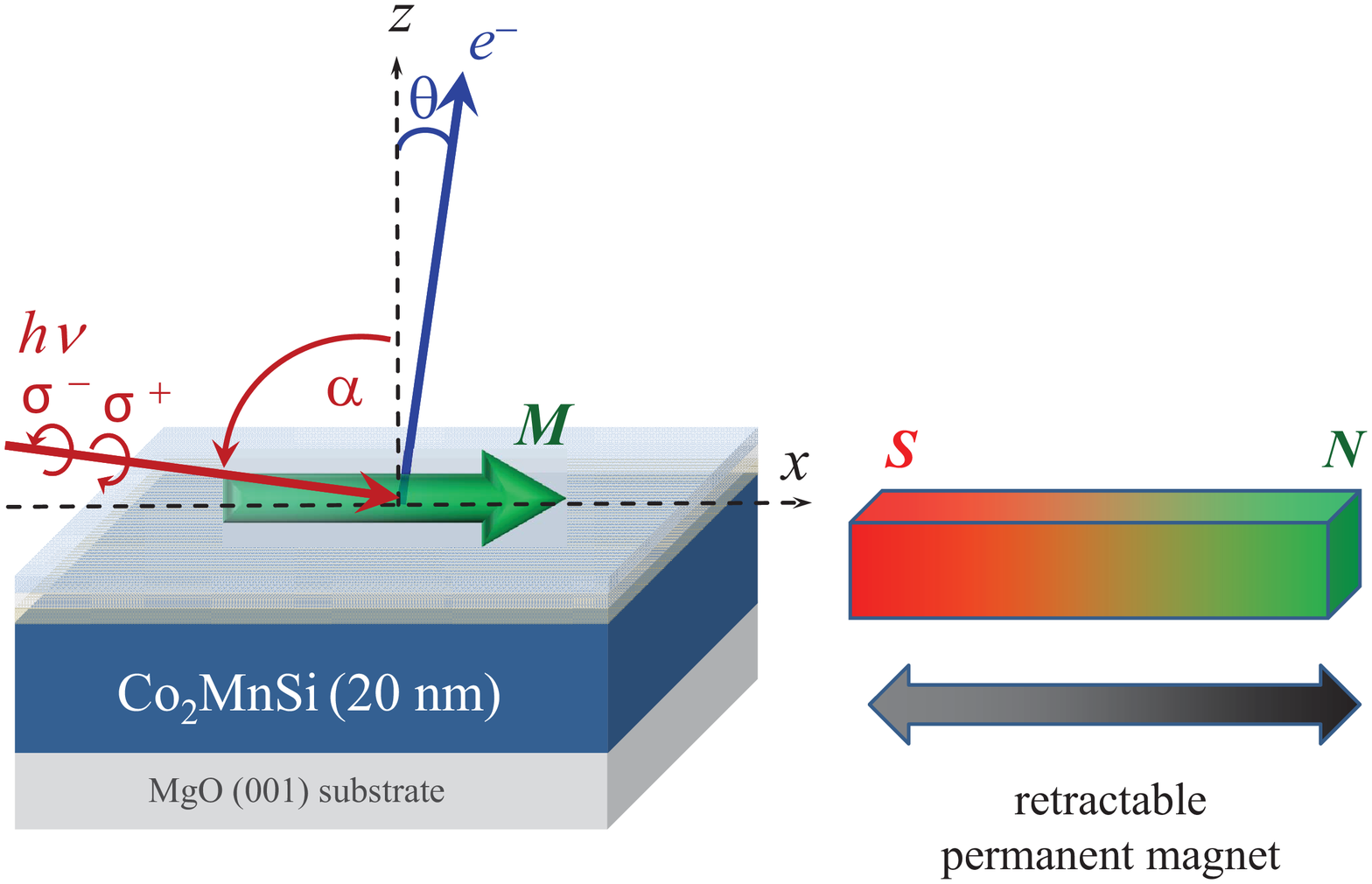,width=8cm}}
   \caption{(Color online) HAXPES setup for MCDAD measurements of Co$_2$MnSi thin films.\\
            The remanent magnetisation $M$ is changed {\it in situ} by a retractable 
            Fe-Nd-B permanent magnet supplying an induction field of about $\pm1$~T.
            The measurements are performed with circularly polarised photons of
            opposite helicity ($\sigma^+$ or $\sigma^-$) that impinge on the 
            sample with an angle of incidence of $\alpha=89^\circ$. The electrons 
            are detected in a near-normal emission geometry with $\theta=1^\circ$ 
            and an angular resolution of about $\Delta\theta\pm15^\circ$. }
\label{fig:setup_mcdad}
\end{figurehere}

The X-ray absorption spectroscopy and X-ray magnetic circular dichroism 
experiments were performed at the undulator beamline ID08 of the European 
Synchrotron Radiation Facility (ESRF, Grenoble, France). The measurements were 
performed in a UHV chamber with a pressure in the low $10^{-10}$~mbar range at 
room temperature. XAS was conducted at the Mn and Co $L_{3,2}$ edges in the 
energy ranges of 625 to 685~eV and 765 to 835~eV, respectively. The energy 
resolution of the {\it Dragon}-type monochromator was set to about 250~meV 
($E/\Delta E > 2.5\times 10^3$). The degree of circular polarisation delivered 
by the {\it Apple II}-type undulator is larger than 99\%. A rapidly switchable 
high-field magnet was used to obtain the XMCD signal. The induction field was 
fixed at $\mu_0H=4$~T and ensures saturation in the out-of-plane geometry. 
Measurements were performed for a {\it normal} geometry with the photon beam 
parallel to the surface normal and for a {\it grazing} geometry with an angle of 
incidence of $\theta=70^\circ$. The magnetic field direction was chosen parallel 
to the incident photon beam in both cases. The setup of the XMCD experiment is 
sketched in Figure~\ref{fig:setup_xmcd}.

\begin{figurehere}
   \centerline{\psfig{file=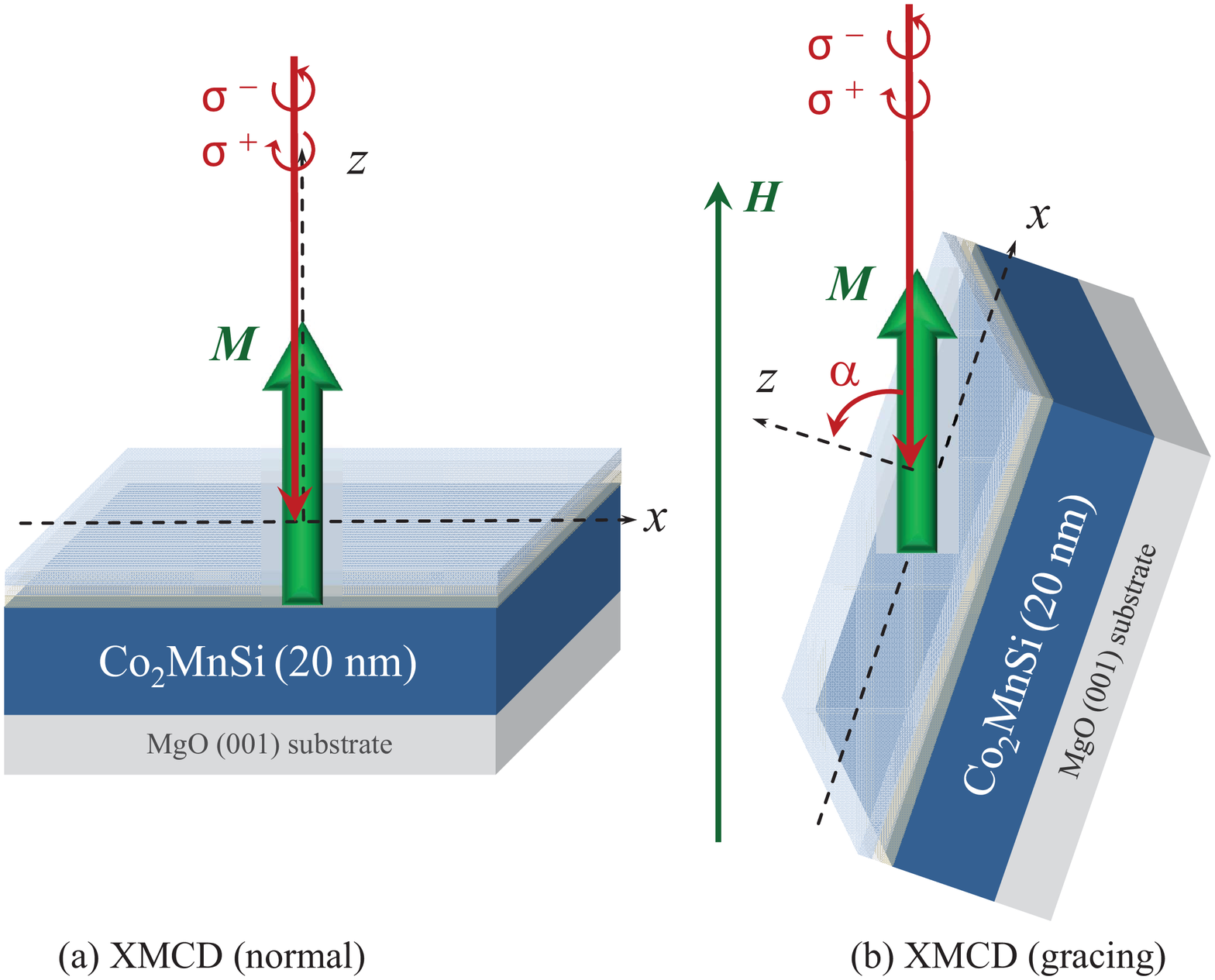,width=8cm}}
   \caption{(Color online) X-ray absorption setup for XMCD experiments on Co$_2$MnSi thin films.\\
            The magnetisation $M$ is induced by a magnetic field $H$ applied
            during the measurement with circularly polarised photons of
            opposite helicity ($\sigma^+$ or $\sigma^-$). Note that
            $H$ and $M$ are generally parallel only in isotropic, soft magnetic materials. }
\label{fig:setup_xmcd}
\end{figurehere}

\section{Results and discussion} 

\subsection{Electronic structure of Co$_2$MnSi} 
\label{sec:elstruc}

The electronic structure was calculated using the full-potential linear 
augmented plane wave (FPLAPW) method as implemented in {\sc Wien}2k~\cite{BSM01,SBl02}.
The details of the calculations are reported in References~\cite{KFF07b,FCF13}.
The atoms were placed on the 8c (Co), 4b (Mn), and 4a (Si) Wyckoff positions of
the $L2_1$ structure with space group $F\:m\overline{3}m$ (225). 
The charge density and other site-specific properties were analysed using 
Baader's quantum theory of atoms in molecules (QTAIM)~\cite{Bad90} using the 
built-in routines of {\sc Wien}2k as well as the {\sc Critic}2 package of the 
programs~\cite{OBP09,OJL14}.

Figure~\ref{fig:dos} shows the result of the electronic structure calculations. 
The compound exhibits a DOS that is typical of HMFs. The minority density has a 
band gap at the Fermi energy. A closer inspection reveals that the size of the 
band gap is determined by states attributed to the Co atoms. The minority gap 
has a width of $\Delta E=0.82$~eV. The top of the minority valence band is about 
0.32~eV below $\epsilon_F$. Owing to the half-metallic character, the magnetic 
moment in the primitive cell is 5~$\mu_B$, as expected from the Slater--Pauling 
rule~\cite{FKW06}.

\begin{figurehere}
   \centerline{\psfig{file=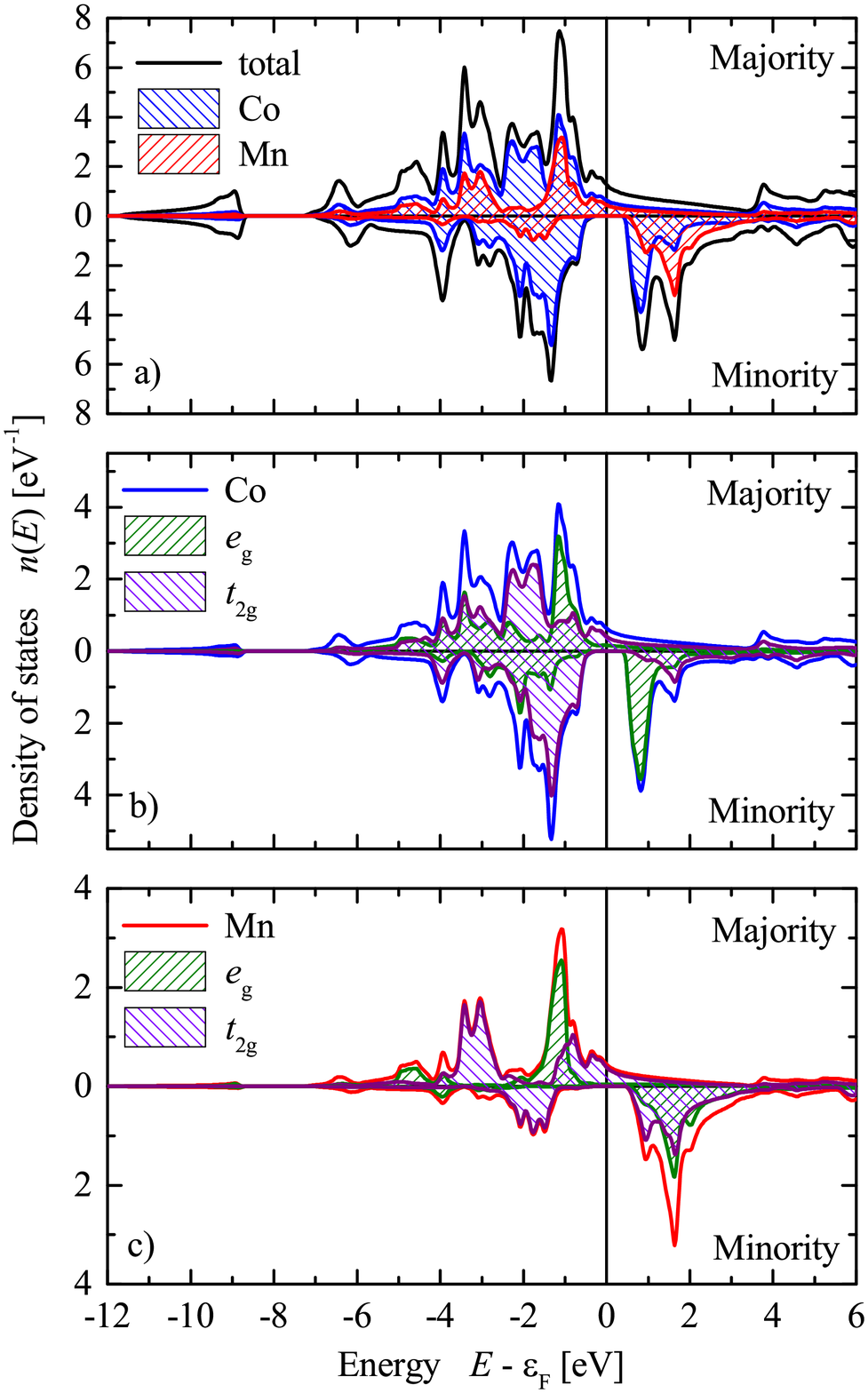,width=6cm}}
   \caption{(Color online) Electronic structure of Co$_2$MnSi. \newline
         (a) Total DOS,
         (b) Co PDOS,
         (c) Mn PDOS.}
\label{fig:dos}
\end{figurehere}

Figure~\ref{fig:charge} shows the valence charge density $\rho(\bm r)$ and the 
magnetisation density $\sigma(\bm r)$ in the (110) plane of Co$_2$MnSi. The magnetisation density 
$\sigma(\bm r)=\rho_\uparrow(\bm r)-\rho_\downarrow(\bm r)$ is calculated from 
the difference of the spin densities arising from majority ($\uparrow$) and 
minority ($\downarrow$) valence electrons, whereas the valence charge density is 
calculated from the sum $\rho(\bm r)=\rho_\uparrow(\bm r)+\rho_\downarrow(\bm r)$. 
The magnetisation density is higher at the Mn atoms compared to the Co atoms and 
vanishes at the Si atoms, as expected. Baader's QTAIM analysis was used to 
analyse the charge density and magnetic moments. The results of the QTAIM 
analysis are listed in Table~\ref{tab:qtaim}. A small charge transfer is 
observed. On average, about 0.3 electrons are transferred from the Mn and Si 
atoms to the Co atoms.

\begin{figurehere}
   \centerline{\psfig{file=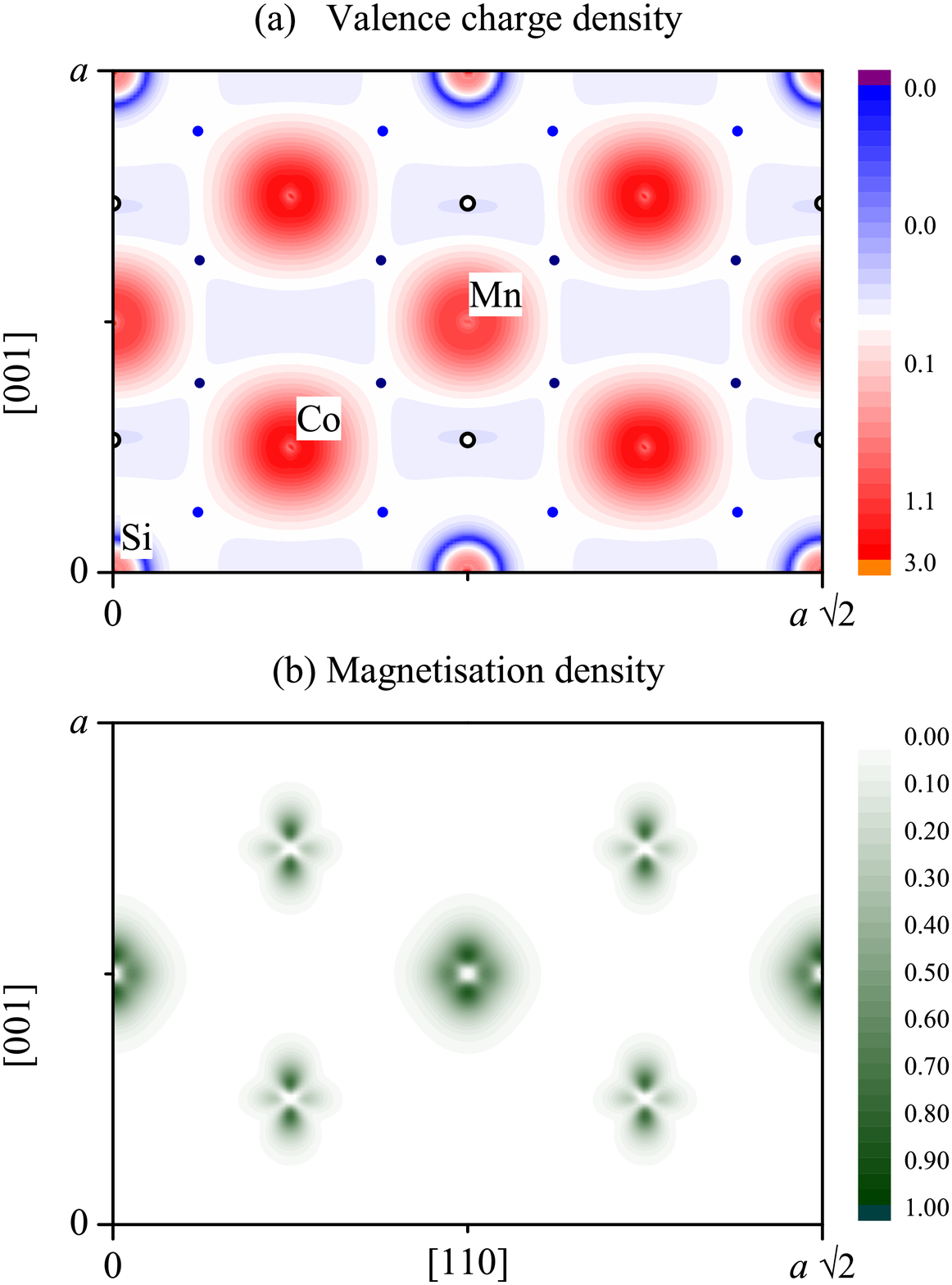,width=8cm}}
   \caption{(Color online) Charge and spin density of Co$_2$MnSi. \newline
         Shown are the valence charge density (a)
         and the magnetisation density (b) in the (110) plane.
         Mn is in the centre of the cell, and Si is located at (0,0).
         The bond and cage critical points are marked by closed and
         open symbols, respectively (see also Table~\ref{tab:crit} for the positions). 
         Note the logarithmic colour scale of the charge density.
         Both densities are given in atomic units.}
\label{fig:charge}
\end{figurehere}

\begin{table}[H]
\tbl{ QTAIM analysis of Co$_2$MnSi. \newline
            $V$ are the basin volumes $V$,
            $n_e$ is the number of electrons in the basin, and $q=Z-n_e$ is the 
            electron excess/deficiency with respect 
            to the electron occupation in free atoms $Z$.  
            Negative values indicate electron excess, that is, negatively charged ions.
            $m$ is the magnetic moment in the basin in multiples 
            of the Bohr magneton. }
      {\begin{tabular}{@{}l cccc@{}}
      \toprule
                     & $V$ [{\AA}$^3$] & $n_e$   &  $q$   & $m$ [$\mu_B$]   \\
      \colrule
      Co             & 11.525          & 27.345  & -0.345 & 1.064  \\
      Mn             & 9.975           & 24.467  &  0.533 & 2.972  \\
      Si             & 11.909          & 13.841  &  0.159 & -0.028 \\
      \botrule   
      \end{tabular} }
\label{tab:qtaim} 
\end{table}

The topology of the charge density $\rho(\bm r)$ is characterised by critical 
points where $\rho(\bm r)$ has an extremal value, that is for $\nabla\rho(\bm 
r)=0$. Besides the gradient $\nabla\rho(\bm r)$, the extrema or critical points 
of $\rho(\bm r)$  are also characterised by the Laplacian $\nabla^2\rho(\bm r)$. 
Four different types of critical points {\bf ({\it r,s})} exist in solids that 
are classified by their rank {\bf \it r} and signature {\bf \it s}:
\begin{itemize}
	\item {\bf (3,-3)} are local maxima of $\rho(\bm r)$. These points appears 
	        at the position of an atom and are called {\it nuclear} critical points. 
  \item {\bf (3,-1)} are saddle points with a local minimum of $\rho(\bm r)$ in 
        1 direction of space and a maximum in the other two. These points appear 
        on the line between two neighbouring atoms and thus define the bond 
        between them. Therefore, they are called {\it bond} critical points. 
  \item {\bf (3,+1)} are also saddle points but with a maximum in one and a 
        minimum in the other two directions of space. These are called {\it 
        ring} critical points because they appear in the middle of several bonds 
        forming a ring.
  \item {\bf (3,+3)} are local minima of $\rho(\bm r)$ and called {\it cage} 
        critical point.
  \item {\bf (0, 0)} is a critical point at infinity that appears only in 
        molecules but not in solids.
\end{itemize}
The QTAIM critical points of Co$_2$MnSi and their properties are summarised in 
Table~\ref{tab:crit}. There are, indeed, three different nuclei that act as 
attractors. A cage critical point $c$ is found between Mn and Si along the [001] 
axis and acts as a repeller. It is an absolute minimum of the charge density. 
Further, two bond critical points $b_{1,2}$ are found; they are located between 
Co and Si ($b_1$) and between Co and Mn ($b_2$). When the two ring critical 
points $r_{1,2}$ are also considered, the
Morse invariant relationships: $n_n>1, n_b>3, n_r>3, n_c>1$ are fulfilled
and the Morse sum of the numbers $n_i$ of the different critical points vanishes 
($n_n-n_b+n_r-n_s=0$)~\cite{ZBa94}, as expected for crystals. The latter is also 
known as the Euler or the Poincar{\'e}--Hopf relation.

\begin{table}[H]
\tbl{ QTAIM critical point analysis of Co$_2$MnSi. \newline
            $pg$ is the point group symmetry of the critical point, and $W$ is 
            the Wyckoff position including the multiplicity of the critical 
            points in the full cubic cell; the multiplicities
            in the primitive cell are one quarter of those values. }
      {\begin{tabular}{@{}ll ccc cr@{}}
      \toprule
      $pg$     & type       &  \multicolumn{3}{c}{position}         &    $W$  &  name     \\    
      \colrule
      $O_h$    & nucleus    &  0          & 0          & 0          &    4a   &  Si       \\    
      $O_h$    & nucleus    &  1/2        & 1/2        & 1/2        &    4b   &  Mn       \\    
      $T_d$    & nucleus    &  1/4        & 1/4        & 1/4        &    8c   &  Co       \\    
      $C_{3v}$ & bond       &  0.1198     & 0.1198     & 0.1198     &   32f   &  $b_1$    \\    
      $C_{3v}$ & bond       &  0.6221     & 0.6221     & 0.6221     &   32f   &  $b_2$    \\    
      $D_{2h}$ & ring       &  0          & 3/4        & 3/4        &   24d   &  $r_1$    \\    
      $C_{2v}$ & ring       &  0          & 0.2049     & 0.2951     &   48i   &  $r_2$    \\    
      $T_d$    & cage       &  0          & 0          & 0.7357     &   24e   &  $c$      \\    
      \botrule
      \end{tabular}}
\label{tab:crit} 
\end{table}

The analysis of the bonding type on hand of the properties of the critical 
points is discussed in Reference~\cite{MPL02}. Metallic systems exhibit a flat 
electron density $\rho$ throughout the valence region. The flatness 
$f=\rho^{c}_{\min}/\rho^{b}_{\max}$ is a measure of the metallicity. 
$\rho^{c}_{\min}$ is the cache critical point, at which the density is minimum, 
and $\rho^{b}_{\max}$ is the highest density among all the bond critical points. 
For Co$_2$MnSi, it is $f = 0.665$. This is of the same order of magnitude as the 
flatness in Cu or Fe (both $\approx0.57$; see Reference~\cite{MPL02}), whereas 
compounds with covalent bonding typically have $f$ values of less than 0.1. From 
the large flatness, the bonding in Co$_2$MnSi is clearly metallic.

Photoabsorption and electron emission spectra have been calculated using 
single-electron as well as many-electron approaches. The single-particle calculations are 
based on the full-potential, fully relativistic spin-polarised {\sc Munich spr-kkr} 
package of Ebert {\it et al.}~\cite{EKM11}. The results of the electronic 
structure calculations are the same as those obtained with the FPLAPW scheme of 
{\sc Wien}2k. In particular, the half-metallic character is retained in the 
fully relativistic calculations, and the semi-relativistic spin-resolved band 
structures and DOS are nearly identical.

The core-levels themselves are strongly localized in the spherical part of the 
potential around the nuclei and behave like atomic states. Atomic-type many-particle
calculations were performed to explain some details of the Mn $2p$ 
states in the photon absorption and electron emission spectra. The multiplet 
calculations were performed using de~Groot's program {\sc ctm4xas}~\cite{SGr10}, 
which includes the effects of the crystal field and charge transfer. The details 
of the multiplet description and applied methods are given in 
References~\cite{Cow81,Gro05,GKo08}. For the calculation of the $2p$ excitation, 
the Slater integrals were scaled to 90\% of their value from the Hartree--Fock 
calculations. The spectra were broadened by 150 to 250~meV according to the 
experimental resolution. The lifetime broadening was varied over the spectrum, 
with larger values used for the "$p_{1/2}$" parts of the spectra to account
for Coster-Kronig contributions.

\subsection{HAXPES} 
\label{sec:haxpes}

The polarisation-dependent core-level spectra near the Co and Mn $2p$ 
excitations are shown in Figures~\ref{fig:co2p} and~\ref{fig:mn2p}, 
respectively. It should be  emphasised that the spectra were taken from 
remanently magnetised samples with no applied external field during the 
measurement. The magnetic moment is thus only about half of the saturation 
moment of 5~$\mu_B$ (see Section~\ref{sec:exp}). The dichroism in the HAXPES 
spectra is here quantified by an asymmetry that is defined as

\begin{equation}
	  A = \frac{I_R-I_L}{max(I_0-I_{\rm bg})}.
\label{eq:asy}	
\end{equation}

$I_R$ and $I_L$ are the intensities for opposite helicity; their difference, 
$I_R-I_L=I_{\rm CD}$, is the dichroism; $I_0=I_R+I_L$ is the sum of the 
intensities; and $I_{\rm bg}$ is the intensity of the background at just above 
the energy of the $2p_{3/2}$ (or $2p_{1/2}$) excitation, where the dichroism is 
zero.

Figure~\ref{fig:co2p}(a) shows $I_0$ and $I_{\rm CD}$ in the energy region of 
the Co $2p$ states. The Co $2p_{3/2}$ excitation appears just above the Mn $2s$ 
state, which complicates the determination of the background in the energy 
region away from the Co $2p$ states. A possible dichroism of the Mn $2s$ state 
is very small and not detectable with the present settings. The Co $2p$ state 
exhibits a spin orbit splitting of $\Delta_{SO}=14.6$~eV into the $2p_{1/2}$ and 
$2p_{3/2}$ substates. $I_{\rm CD}$ exhibits a change of sign in the series $+--+$ 
across the energy range of the $2p$ excitation. This is typical of a Zeeman-type 
level ordering in the single-electron model~\cite{Men98}. A pronounced 
satellite is observed at about 4.3~eV below the $2p_{3/2}$ state but is not 
detectable at the $2p_{1/2}$ state. The polarisation-dependent spectra and 
dichroism near the $2p_{3/2}$ state are shown in Figure~\ref{fig:co2p}(b). The 
satellite obviously exhibits negative dichroism. Further, the polarisation-
dependent spectra reveal that the main $2p_{3/2}$ excitation exhibits a 
splitting of about 200~meV. At the same time, the asymmetry, as defined by 
Equation~(\ref{eq:asy}), varies between +17\% and -8\% across the $2p_{3/2}$ 
part and between -34\% and +25\% across the $2p_{1/2}$ part of the spectra. Both 
the polarisation-dependent spectra and the dichroism indicate that the lines of 
the multiplet extend over the entire spectral range. In particular, the 
dichroism does not vanish between the two main parts of the spin-orbit doublet.

\begin{figurehere}
   \centerline{\psfig{file=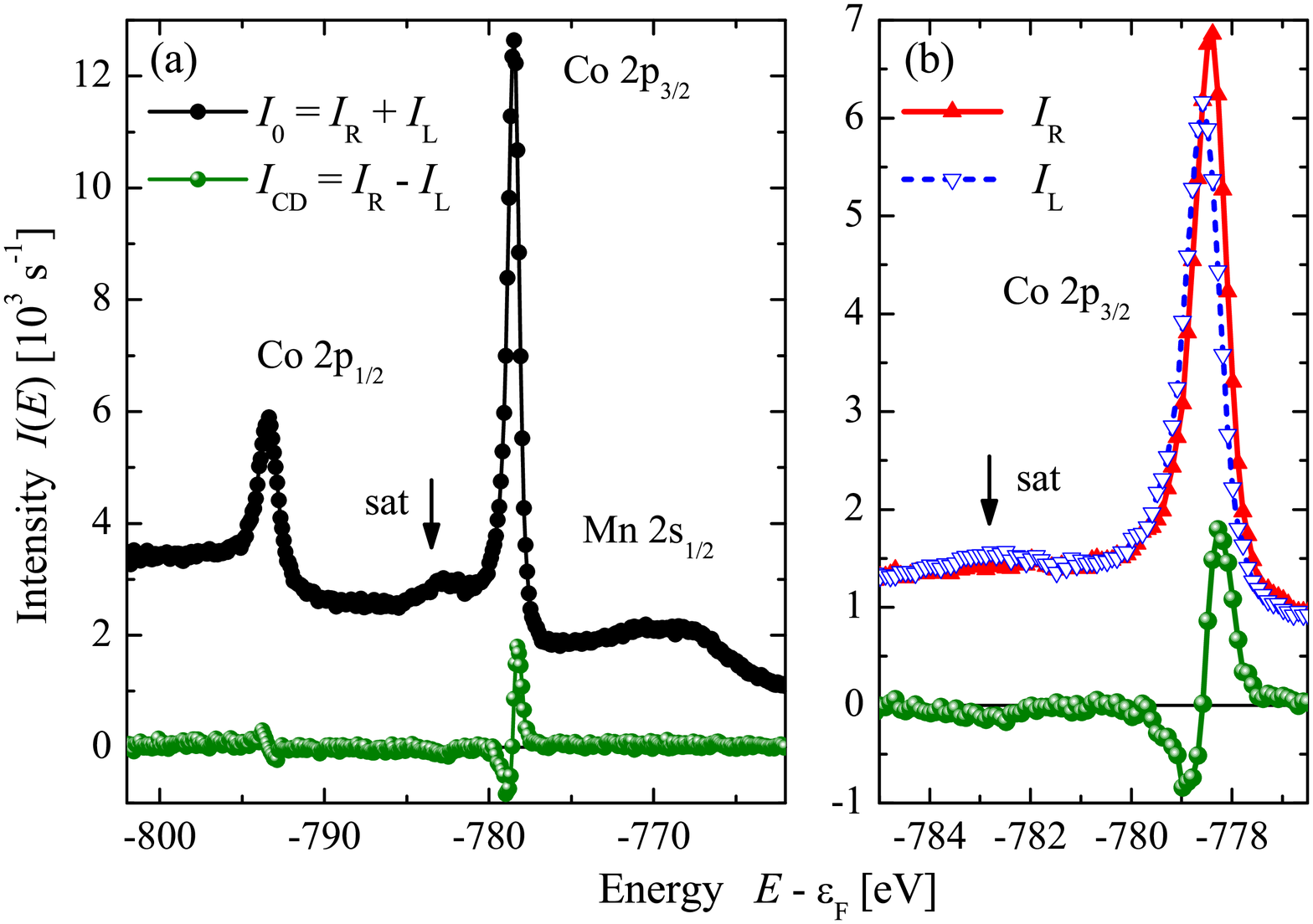,width=8cm}}
   \caption{(Color online) Co $2p$ and Mn $2s$ HAXPES spectra of Co$_2$MnSi on MgO(100). \newline
            (a) Sum and difference of spectra taken with photons of opposite 
                helicity, 
            (b) spectra at $2p_{3/2}$ taken with $\sigma^+$ (R) 
                and $\sigma^-$ (L) polarisation of the photons.
            Arrows indicate the {\it "4~eV"} satellite. } 
\label{fig:co2p}
\end{figurehere}

Figure~\ref{fig:co2pcalc} compares Co $2p$ photoelectron spectra that were 
calculated using different schemes. In all the calculations, a fixed lifetime 
broadening was assumed for the complete spectrum to make the single particle and 
many electron calculations comparable. Therefore, the details of the intensity 
may differ between the experiment and calculations. Note further that the energy 
scale of the single electron calculations in Figure~\ref{fig:co2pcalc}(a) is 
related to the ground-state binding energies. Better absolute values for the 
energies may be found by introducing core holes and using Slater's transition 
state theory.

\begin{figurehere}
   \centerline{\psfig{file=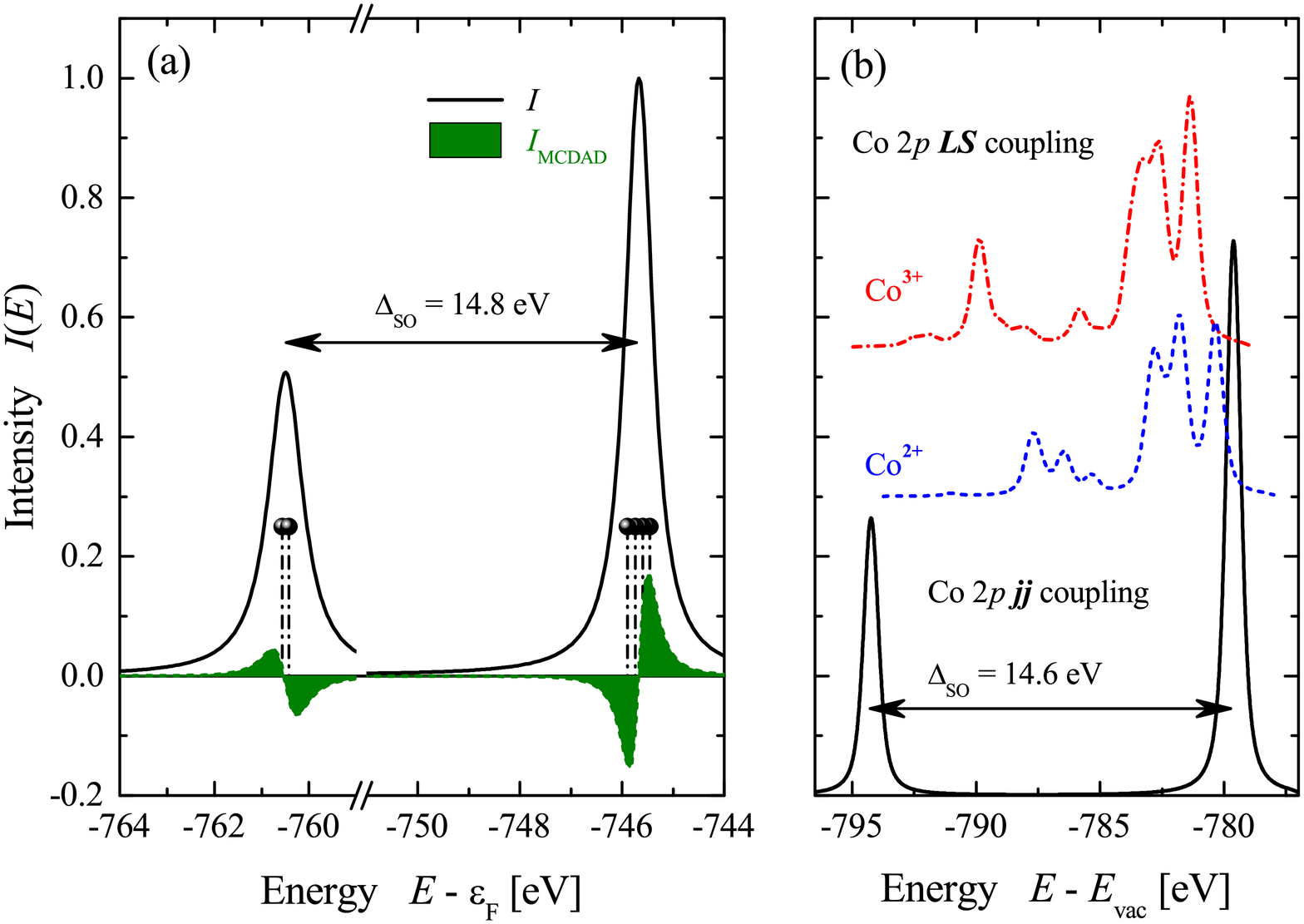,width=8cm}}
   \caption{(Color online) Calculated Co $2p$ photoelectron spectra of Co$_2$MnSi. \newline
            (a) Spectrum and dichroism calculated by {\sc spr-kkr} for the solid state 
                (note the $x$ axis break), 
            (b) spectra calculated by {\scshape ctm4xas} for atoms, assuming different
                coupling schemes ($jj$ and $LS$) and states ($2+$ and $3+$) of Co. } 
\label{fig:co2pcalc}
\end{figurehere}

Figure~\ref{fig:co2pcalc}(a) shows the results of the relativistic {\sc spr-kkr} 
formalism using the full solid-state potential, whereas 
Figure~\ref{fig:co2pcalc}(b) shows the results of atomic-type Hartree--Fock 
calculations with {\scshape ctm4xas}. The atomic-type calculations were 
performed assuming different states of Co: Co$^{2+}$ and Co$^{3+}$. The spectra 
shown in Figure~\ref{fig:co2pcalc}(b) are for pure $jj$ and $LS$ couplings. The 
spectra for $LS$ coupling show typical multiplet structures. Those structures, 
however, do not appear in the experimental spectra, which show only the spin-orbit 
doublet and a single satellite. Calculations for intermediate couplings 
($LSJ$) did not improve the situation. This observation is in accordance with 
multiplet calculations reported by other authors~\cite{MGR06}.

The splitting of the $2p$ state in $jj$ coupling agrees well with the spin-orbit 
splitting found in the full relativistic {\sc spr-kkr} calculations. The main 
features of the spectra in the {\sc spr-kkr} calculations agree well with the 
measured spectra, in particular the $+--+$ series of the sign of the dichroism. 
The {\it 4 eV} satellite is not explained by the solid-state calculations. Note 
that those calculations did not include inelastic or many-body effects. The 
$2p_{1/2}$ and $2p_{3/2}$ lines are split by the magnetic interactions according 
to their $m_j$ substates. The positions of the sublevels are plotted with 
vertical lines in Figure~\ref{fig:co2pcalc}(a). The exchange interaction results 
in Zeeman-type splitting with $\Delta_{\rm exc}\approx 150$~meV between 
neighbouring $m_j$ levels.

The $2p$ spectra of Mn, as shown in Figure~\ref{fig:mn2p}, exhibit a more 
complicated structure than the Co $2p$ core level. Splittings of 
$\Delta_{1/2}=0.7$~eV and $\Delta_{3/2}=1.2$~eV are revealed at the $2p_{1/2}$ 
and $2p_{3/2}$ excitations, respectively. The plot of the total intensity $I_0$ 
in Figure~\ref{fig:mn2p}(b) obviously does not unambiguously reveal a spin-orbit 
splitting due to the additional splitting of both lines, $2p_{3/2}$ and 
$2p_{1/2}$. The mean splitting between the doublet-type structure amounts to 
about $\Delta=11$~eV. The asymmetry across the $2p_{3/2}$-type part of the 
spectrum varies between +48\% and -16\%. The dichroism does not vanish between 
the $2p_{1/2}$ and $2p_{3/2}$ lines. This confirms that the splitting is not of 
Zeemann type, in that case no additional states would appear between the main 
lines of the spin-orbit doublet.

\begin{figurehere}
   \centerline{\psfig{file=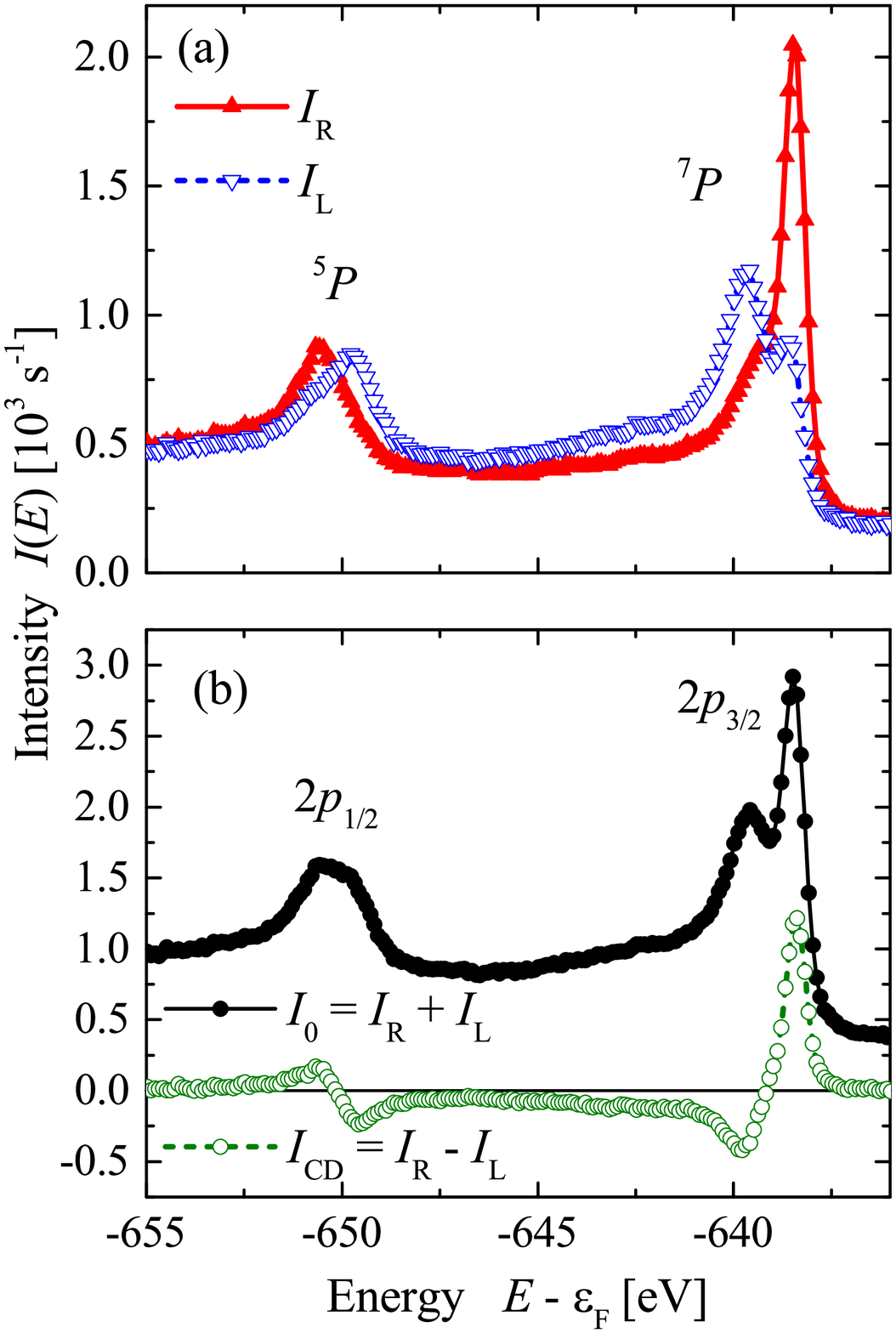,width=6cm}}
   \caption{ (Color online) Mn $2p$ HAXPES spectra of Co$_2$MnSi on MgO(100). \newline
             (a) Spectra taken with $\sigma^+$ (R) and $\sigma^-$ (L) 
                 polarisation of the photons,
             (b) sum and difference of spectra taken with photons of opposite 
                 helicity. }
\label{fig:mn2p}
\end{figurehere}

The bare intensity spectra ($I_0$) of the Co and Mn $2p$ states agree well with 
spectra taken with linearly polarised photons from bulk material of the 
isovalent compound Co$_2$MnGe~\cite{OFB11}. The dichroism at the Co $2p$ states 
is close to that observed for exchange-biased CoFe or Co$_2$FeAl 
films~\cite{KFS11}. Here, the asymmetry is lower because of the relaxed 
resolution used in the present work. Further details of the spectra and 
dichroism will be discussed after the soft X-ray absorption spectra of the 
$L_{2,3}$ edges of Mn and Co are presented.

\subsection{XAS, XMCD} 
\label{sec:xas}

Figure~\ref{fig:xmcd} shows the X-ray absorption spectra and magnetic circular 
dichroism taken at the $L_{2,3}$ edges of Mn (a) and Co (b). The spectra agree 
well with those reported by other 
authors~\cite{SKS04,SMH05,TKL06,SKI07,SKE08,TKL08,TKS08}. The striking 
differences in the HAXPES spectra are easily recognised. A splitting of the 
$L_3$ edge (corresponding to the $2p_{3/2}$ excitation) is not detectable for 
either Mn or Co. The spin-orbit splitting of the Co $2p$ states measured between 
the $L_3$ and $L_2$ maxima amounts to 15.2~eV; thus, it is slightly larger than 
the value observed in the electron emission spectra. A satellite is observed 
about 3.5~eV above the $L_3$ white line of Co. It does not exhibit any 
pronounced dichroism and thus should have a different origin from the 4.3~eV 
satellite observed in electron emission. It might, therefore, not belong to a 
multiplet-type splitting. Other than in the HAXPES-MCD where the asymmetry 
changes sign within each line of the doublet, the soft XMCD (SXMCD) of 
Co does not exhibit a change of sign across the $L_3$ or $L_2$ absorption edges. 
The sign of the SXMCD changes only between the two lines of the spectra.
It is opposite at the two different edges and the change of sign takes place
at 9.8~eV above the $L_3$ line. 

The asymmetry of the XMCD is defined by

\begin{equation}
	  A = \frac{I^+-I^-}{I^++I^-},
\label{eq:asyxmcd}	
\end{equation}

where $I^\pm$ are the intensities for photon helicity parallel (+) and 
antiparallel (-) to the magnetic field after background subtraction. At the 
$L_3$ edge of Co, it amounts to -33\% and has approximately half that value at 
the $L_2$ edge.

\begin{figurehere}
\centering
   \centerline{\psfig{file=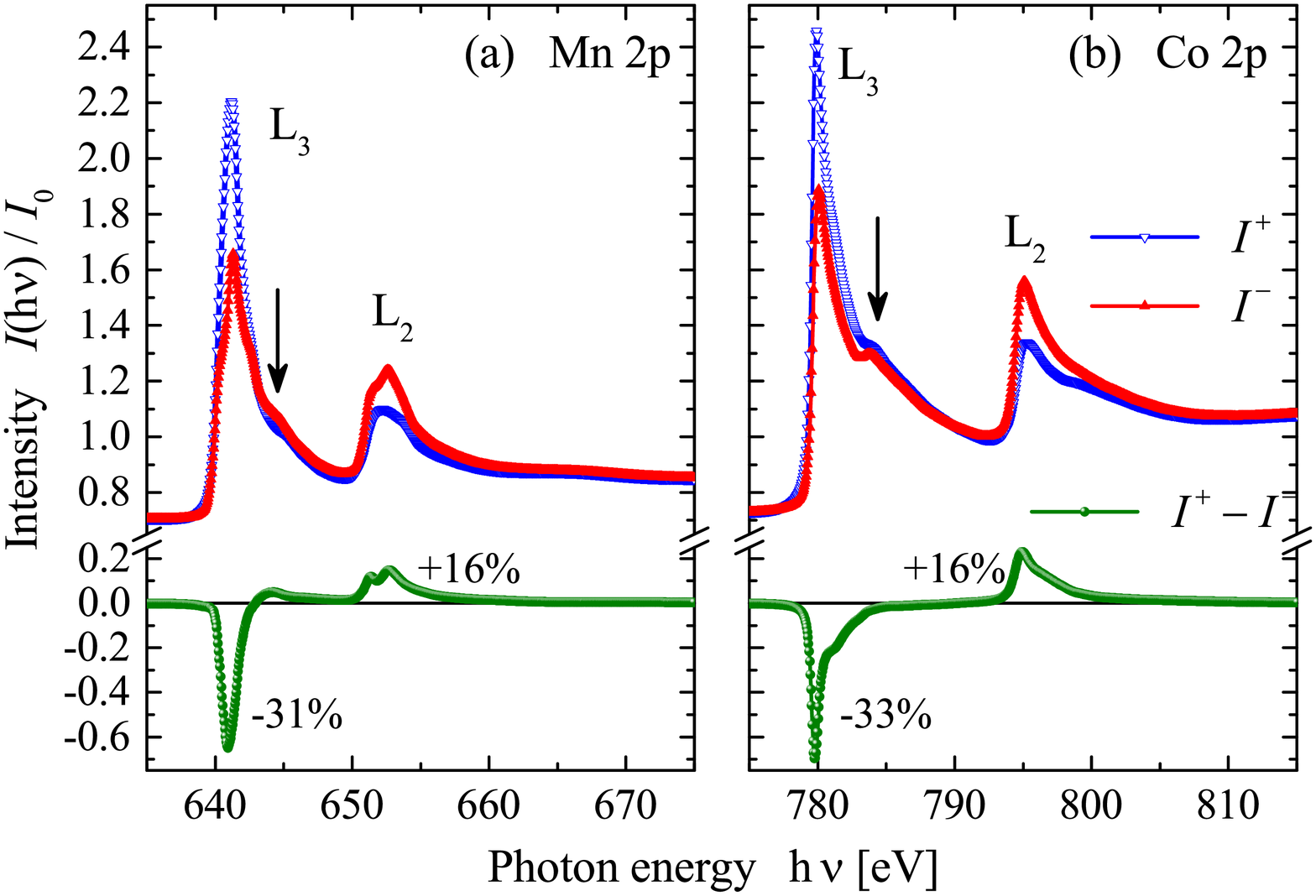,width=8cm}}
   \caption{ (Color online) Co and Mn $2p$ soft X-ray absorption spectra of Co$_2$MnSi on MgO(100). \newline
             (a) Spectra and dichroism taken at the Mn $L_{2,3}$ white line,
             (b) spectra and dichroism taken at the Co $L_{2,3}$ white line.
             The intensity is normalised by the photon flux. }
\label{fig:xmcd}
\end{figurehere}

The Co $L_{3,2}$ spectrum was calculated with {\sc spr-kkr} and is shown in 
Figure~\ref{fig:Coxmcdcalc}. The calculated spectrum contains the {\it 4~eV} 
satellite that appears in the measured spectra. This satellite was previously 
also observed in other Heusler compounds based on 
Co$_2$~\cite{EFV03,EWF04a,WFK05,WFK06a,OFB11}. It is due to a transition into 
empty states with high density. It is seen at about 4~eV above the Fermi energy 
in the majority density of Co, as plotted in Figure~\ref{fig:dos}(b), and also 
in the unoccupied part of the total majority DOS. Those empty states  in 
photoelectron spectra may also serve as final states for interband transitions 
from states at the Fermi energy. Also shown in Figure~\ref{fig:Coxmcdcalc}(b,c) 
are the differential spin and orbital moments that fulfill the differential sum 
rules~\cite{Ebe96}. 

\begin{figurehere}
\centering
   \centerline{\psfig{file=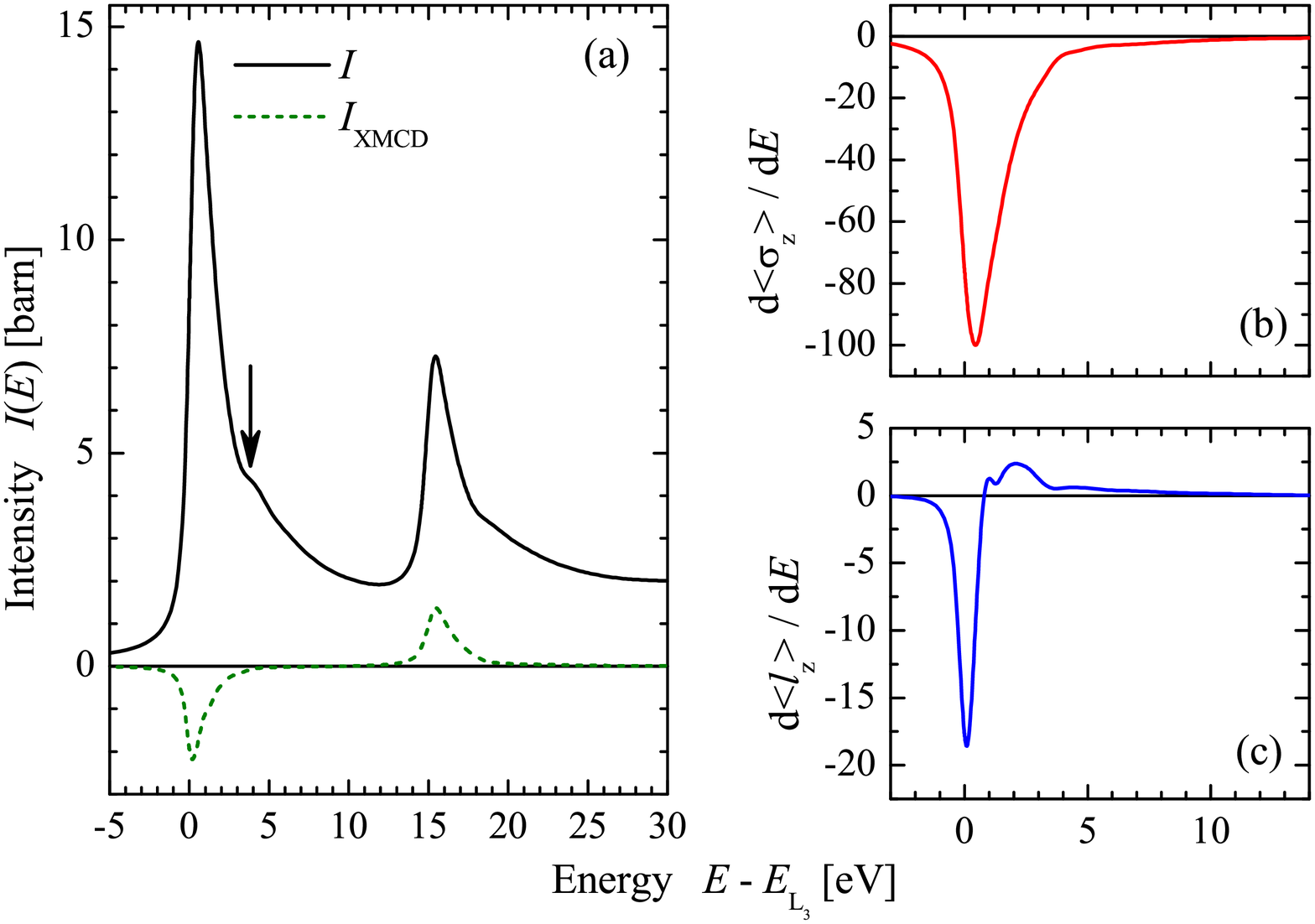,width=8cm}}
   \caption{ (Color online) Calculated Co $L_{3,2}$ soft XAS spectrum and XMCD of Co$_2$MnSi. \newline
             (a) Spectrum and dichroism,
             calculated (b) differential spin $\left< \sigma_z \right>$, and 
                 (c) orbital $\left< l_z \right>$ momentum expectation values.
              The arrow in (a) marks the{\it 4~eV} satellite.}
\label{fig:Coxmcdcalc}
\end{figurehere}

Like the Mn $2p$ electron emission spectra in Figure~\ref{fig:mn2p}, the 
absorption spectrum shown in Figure~\ref{fig:xmcd}(a) exhibits several 
satellites. The appearance of metallic satellites or multiplet splittings from 
the exchange interaction is typical of electron emission. The metallic 
satellites arise from plasmon losses or excitation of interband transitions, as 
explained for the Co $2p$ spectra. Most striking in the Mn XAS spectra is that 
the 1.3~eV splitting of the $2p_{3/2}$ state in the HAXPES spectra is not 
observed at the $L_3$ absorption edge. The splitting revealed by two maxima in 
the SXMCD at the $L_2$ edge is 1.3~eV, which is larger than that observed from 
the maxima at $2p_{1/2}$ in the electron emission spectrum. It is, however, of 
the same order as the energy difference of 1.1~eV observed between the negative 
and positive dichroism maxima in the HAXPES-MCD. The asymmetry at the $L_3$ edge 
amounts to -31\%. The XMCD changes sign at 1.6~eV above the $L_3$ maximum. The 
dichroism does not vanish between the $L_3$ and $L_2$ lines, similar to the 
observation in the electron emission spectra. 

In atoms, multiplet splitting is due to the interaction of the $nl^{-1}$ core 
hole with the polarised open valence shell. The core hole (here $2p^5$) in 
solids is expected to interact with the polarised $d$ states of the valence 
band. The localised valence $d$ states, however, are screened by delocalised 
electrons. The well-known multiplet theory may be used to explain the observed 
splittings in the spectra~\cite{Cow81,vdL91,Gro05,GKo08,BBd00}, assuming that 
the atomic character of the valence electrons is partially retained in the 
solid. It was shown above that this approach is critical for Co because of the 
rather delocalised character of the valence $d$ electrons. Therefore, we will 
focus on the Mn atoms in Co$_2$MnSi. The description becomes complicated, as it 
is not {\it a priori} clear what ionic state the Mn adopts in the metal. From 
the electronic structure calculations, one has Mn $d^5$, neglecting all the 
other shells. However, some of the $d$ electrons may be delocalised in the metal 
and do not contribute to the coupling.

For the multiplet analysis, a Mn$^{2+}$ ionic state with a $^6S_{5/2}$ ground 
state in {\it LSJ} coupling is assumed (see~\cite{OFB11} for Mn$^{3+}$). Note 
that the description of the ground states of neutral Mn$^{0}(4s^23d^5)$ and 
Mn$^{2+}(4s^03d^5)$ are principally the same because the filled $4s^2$ shell 
contributes only $^1S_{0}$. According to the dipole selection rules, the 
following transitions are allowed for the  excited states during excitation of 
the $2p$ core level from $2p^63d^5 \: (^6S_{5/2})$:

\begin{itemize}
\item in electron emission to \\
      $ \left\{ \left[ 2p^53d^5 \: (^{5,7}P_{J}) \right] + \epsilon(s,d) \right\} (^6P_{7/2,5/2,3/2}) $,\\
\item or in photon absorption to \\
      $ \left[ 2p^53d^6 \: (^{6}P_{7/2,5/2,3/2}) \right] $ and \\
      $ \left[ 2p^53d^5s^1 \: (^{6}P_{7/2,5/2,3/2}) \right] $.
\end{itemize}

The transition schemes given above contain only the leading terms that can be 
reached according to the selection rules. The final-state $p^5d^6$ configuration 
in the absorption scheme contains 1260 terms with a degeneracy of 180, and 
overall 110 different transitions are allowed by the selection rules, $\Delta J=0, \pm1$. 
In the electron emission scheme, $\epsilon(s,d)$ denotes the ejected 
electrons with kinetic energy $\epsilon_k$ and orbital angular momentum $l'=0,2$ 
for ionisation of the $2p$ shell with $l=1$. The intermediate ionic states in 
electron emission obviously have spin values of $S=5, 7$; thus, the splitting 
observed in electron emission from the $2p$ core level may be assigned to the 
spin-exchange splitting of the ionic states. The smaller splitting observed in 
electron emission from the $2p$ shell corresponds to states with different total 
angular momentum $J$ of the $^7P_{J=4,3,2}$ and $^5P_{J=3,2,1}$ intermediate 
ionic states. The $^7P$ part of the multiplet can be formed only from 
$2p^5(^2P)$ and $3d^5(^6S)$ subshell couplings~\cite{BBJ00}. In particular, the 
$^7P_{4}$ state has the highest energy in the electron emission spectrum. It is 
directly related to the $p_{3/2}$ single particle sub-state of a $(jj)_J$ 
coupling scheme

The result of the multiplet calculations for photoexcitation of the $2p$ core 
level of Mn$^{2+}$ are shown in Figure~\ref{fig:mn2pcalc}. The line spectra as 
well as the broadened emission and absorption spectra are shown. The above-mentioned
manifold of allowed transitions makes the spectra very complex. Many 
details of the line spectra are covered, however, by the lifetime broadening.

\begin{figurehere}
   \centerline{\psfig{file=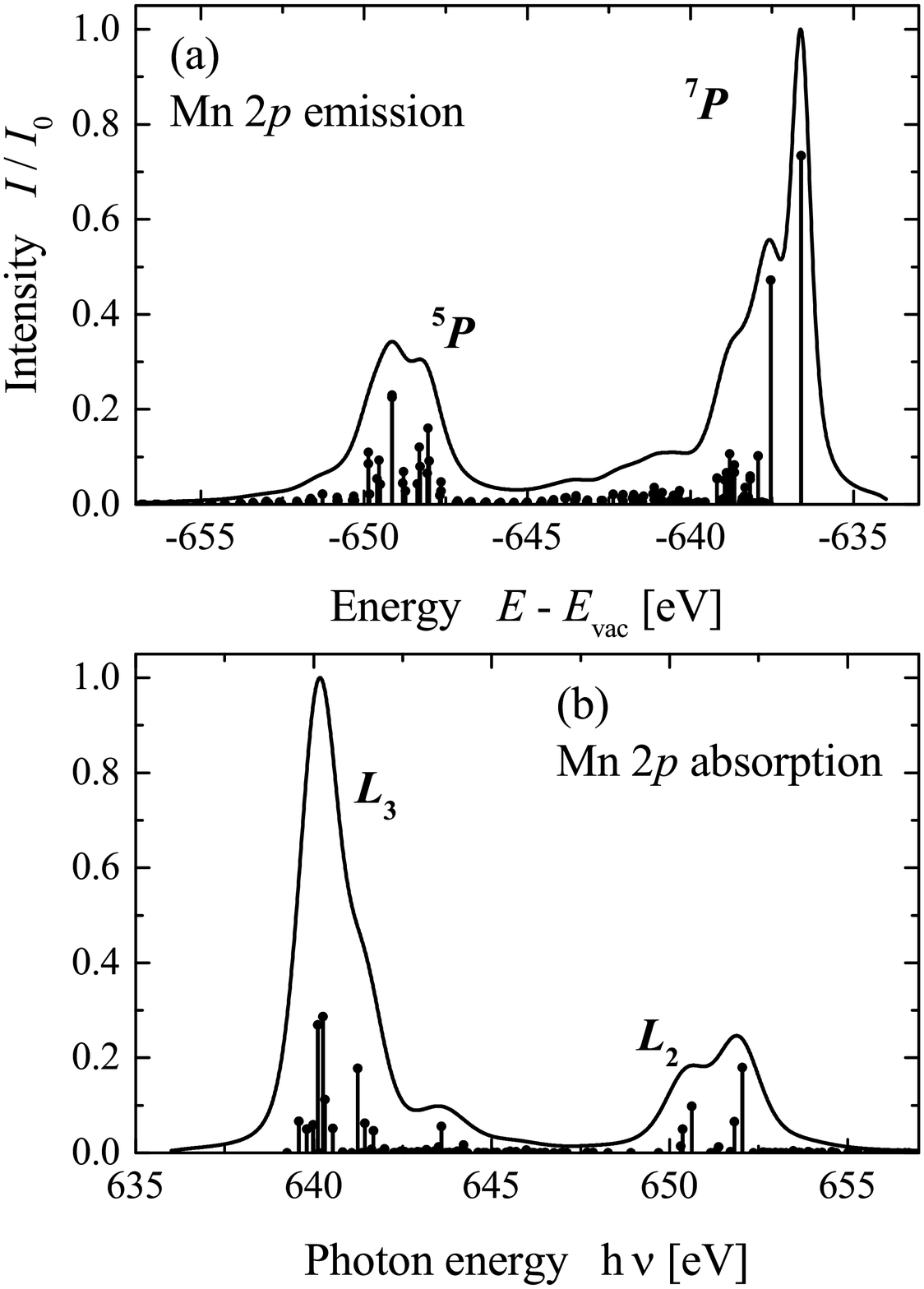,width=6cm}}
   \caption{ Calculated Mn $2p$ XPS and XAS spectra. \\
             (a) Electron emission spectrum,
             (b) photoabsorption spectrum.}
\label{fig:mn2pcalc}
\end{figurehere}

The calculated electron emission spectrum is shown in 
Figure~\ref{fig:mn2pcalc}(a). It was calculated with a crystal field parameter 
of $10D_q=1.5$~eV for the initial state of Mn$^{2+}$. The spectrum exhibits 
splitting at the $^7P$ ($2p_{3/2}$) and $^5P$ ($2p_{1/2}$) lines. The $^7P_4$ 
intermediate ionic state is clearly split off from the remaining spectrum. The 
multiplet lines related to the $^7P$ part of the spectrum extend over a range of 
10~eV. Figure~\ref{fig:mn2pcalc}(b) shows the photoabsorption spectrum. It was 
calculated for the $^6S_{5/2}$ initial 
state of Mn$^{2+}$. The general shape of the spectrum corresponds to that of the 
measurement. The $L_2$ white line exhibits a splitting similar to that observed 
in the experiment. A pronounced feature appears at about 3.5 eV above the 
maximum of the $L_3$ white line. Unlike that in the Co spectrum, this satellite 
is attributed to a multiplet effect. 

The further analysis of the XAS and XMCD data is given in the following using 
the measurements at the Co $L_{3,2}$ edges as an example. The starting point is 
the data normalised to the photon intensity. The {\it `unpolarised'} total 
intensity $I_{\Sigma}=(I^++I^-)/2$ is found from the sum of the intensities for 
the circular polarisation parallel ($I^+$) and antiparallel ($I^-$) to the 
direction of magnetisation. Further, a constant background signal is subtracted 
such that $I_0=I_{\Sigma}-I_{\rm bg}$. The XMCD signal is given by the 
difference $I_{\rm MCD}=(I^+-I^-)$. The difference does not contain any 
background; therefore, it can easily be used to reconstruct the polarisation-dependent 
spectra from $I^\pm=I_0 \pm I_{\rm MCD}/2$. The total intensity $I_0$ 
and the XMCD signal $I_{\rm MCD}$ are plotted in Figures~\ref{fig:xmcdana}~(a) 
and~(b), respectively.

\begin{figurehere}
   \centerline{\psfig{file=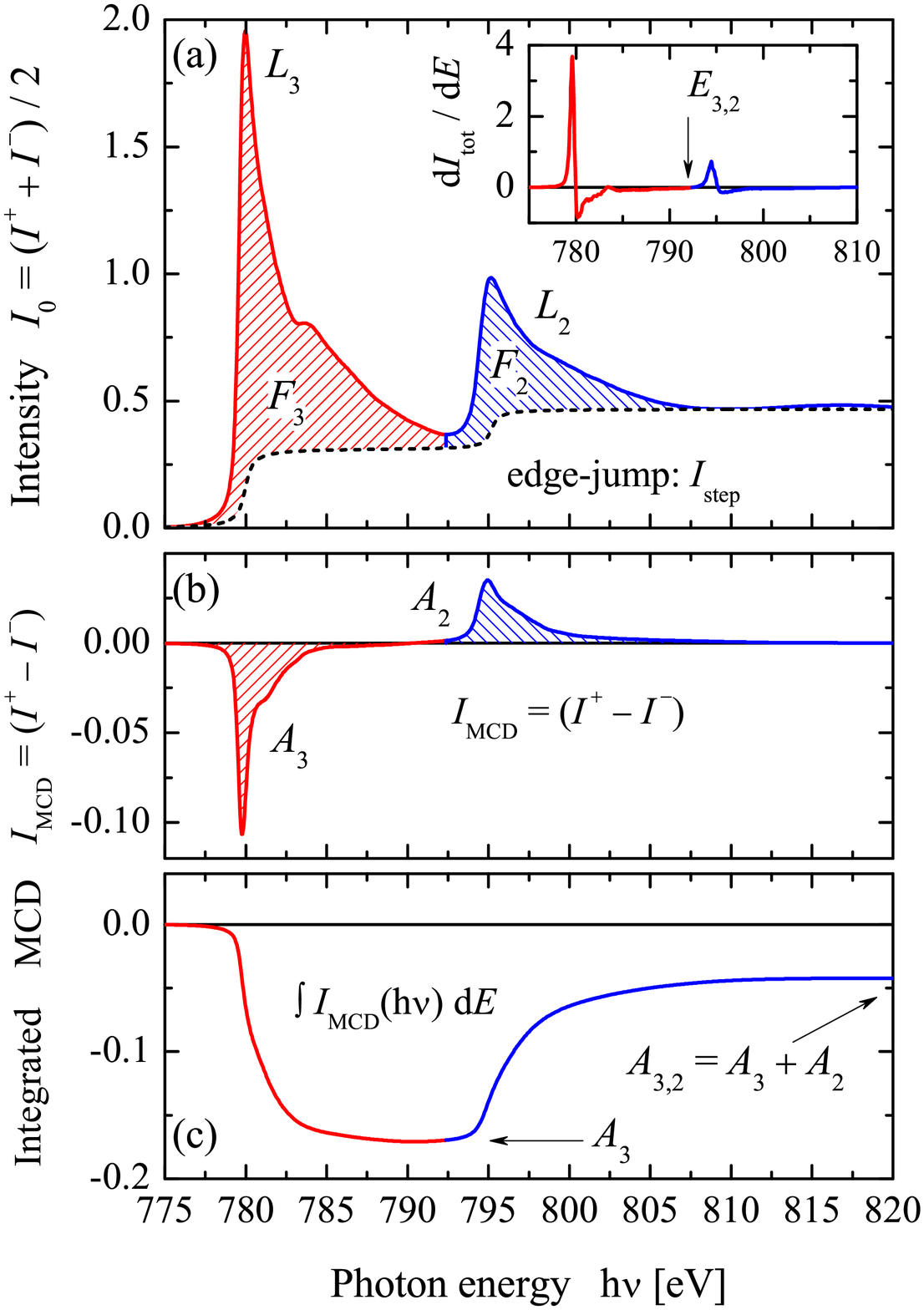,width=6cm}}
   \caption{ (Color online) Analysis of Co XAS and XMCD spectra. \\
             (a) XAS spectrum and edge-jump function
                 after a constant background is subtracted
                [compare Figure~\ref{fig:xmcd}~(b)],
             (b) XMCD signal,
             (c) integrated dichroism.
             Inset in (a) shows the derivative of the spectrum.}
\label{fig:xmcdana}
\end{figurehere}

Several energies have to be extracted from the spectra for further analysis, 
also in view on sum rules. The differentiated total intensity~-- shown in 
the inset of Figure~\ref{fig:xmcdana}~(a)~-- is used for this purpose. The 
$L_{3,2}$ absorption edges are identified by the maxima of the derivative, and the 
energies of the maxima of the corresponding white lines are identified by the nearby zero 
crossings. The energy $E_{3,2}$ for the separation of the $L_3$ from the $L_2$ 
part of the spectrum is taken from the zero crossing of the derivative just in 
front of the $L_2$ white line (see arrow in the inset).

The final normalisation of the intensity and XMCD is performed with respect to 
the white line intensities after an $\arctan$-type step function, 
$I_{step}$, is subtracted for the edge jumps. The step function is modelled by

\begin{eqnarray}
   I_{step} & = &   \frac{h}{3} \left[ 1 + \frac{2}{\pi} \arctan \left( \frac{E-E_3}{w_3} \right) \right ] \nonumber \\
            &   & + \frac{h}{6} \left[ 1 + \frac{2}{\pi} \arctan \left( \frac{E-E_2}{w_2} \right) \right ],  
\label{eq:istep}
\end{eqnarray}

where $h$ is the height and $w_3$ and $w_2$ are the widths of the step function. 
The latter were set to $w=350$~meV for both edges. The used prefactors imply a 
branching ratio of $I_3/I_2=2$ for excitation from the $p$ states into 
delocalised $s$ or $d$ states. The energies $E_3$ and $E_2$ for the steps were 
taken from the maxima of the derivative. In the literature, those energies were 
sometimes taken from the maxima of the white lines. The width $w$ as well as the 
step energies $E_3$ and $E_2$ influence the quantities that are calculated from 
the spectra. In this work, tests indicated that variations in the energies and 
width do not influence the spin magnetic moments by more than 2.5\% for Co or 
4\% for Mn. These variations may be considered small compared to the influence 
of any variation of the photon polarisation or deviations resulting from the 
uncertain number of unoccupied $d$ states or other experimental uncertainties. 

The area between $I_0$ and $I_{step}$ is proportional to the number of 
unoccupied $d$ states and is calculated by integration of 
$I(E)=I_0(E)-I_{step}(E)$.

\begin{eqnarray}
   F & = & F_3 + F_2                         \nonumber \\                            
     & = & \int_{E<E_3}^{E_{3,2}} I(E)dE
            + \int_{E_{3,2}}^{E>E_2} I(E)dE
\label{eq:areas}
\end{eqnarray}

The piecewise integration of $F$ enables calculation of the branching ratio 
$b_i=F_3/F_2$. Note that the branching ratio is not consistently defined in the 
literature. In other work, it may be given as the inverse value 
$r_i=F_2/F_3$~\cite{Goe05} or by $F_3/F$~\cite{GBN05a}. 
Figure~\ref{fig:xmcdana}~(a) shows that $I(E_{3,2})>0$ does not vanish at the 
separation energy $E_{3,2}$, and some intensity of the $2p_{3/2}$ excitation is 
also found below $2p_{1/2}$. This effect leads to a small underestimation of the 
branching ratio when $b_i$ is used.

The ratio $q=F_i/h$ of the areas $F_i$ and the height of the step function is a 
measure of the relative contribution of transitions into localised $d$ states 
to the absorption spectra~\cite{EFV03}. From the data shown in Figure~\ref{fig:xmcd}, one 
finds $q_{\rm Co}=13.2$~eV and $q_{\rm Mn}=22$~eV for Co and Mn, respectively. 
Similarly, from the white line maxima one finds $Q_m=I_{max,L_i}/h$ values of 3.85 
and 7.1. Both ratios are larger by a factor of about 2 for Mn than for Co. 
This hints directly at the stronger localisation of the Mn $d$ valence states.

The areas $A_3$ and $A_2$ underneath the XMCD signal [see 
Figure~\ref{fig:xmcdana}~(b,c)] are needed for the sum rule analysis. They are 
found by integrating $I_{\rm MCD}$. Similar to the areas $F_i$ under the 
intensity, the areas $A_i$ are found by piecewise integration. $A_3$ is found 
from integration up to $E_{3,2}$, and the total integral yields 
$|A_{3,2}|=|A_3|-|A_2|$ because of the opposite sign of the magnetic circular 
dichroism at the two white lines. 

From the sum rule analysis (see References~\cite{CTA93,CIL95}), the spin ($m_s$) 
and orbital ($m_l$) magnetic moments are given by

\begin{eqnarray}
   m_s + m_t & = & \frac{n^+_d}{P_c \cos(\theta)} \: \frac{A_3-2A_2}{F} \: \mu_B,     \nonumber \\
   m_l       & = &  \frac{n^+_d}{P_c \cos(\theta)} \: \frac{2}{3} \: \frac{A_3+A_2}{F} \: \mu_B,
\label{eq:sumrule}   
\end{eqnarray}

where $P_c$ is the degree of circular polarisation, $\theta$ is the angle 
between the magnetisation and the photon momentum, and $n^+_d$ is the number of 
unoccupied $3d$ states. The factor $n^+_d/F$ is used to normalise the data. 
Further, $m_t$ is the magnetic dipole or spin anisotropy term. These equations 
neglect the effect of different matrix elements. (Please note that the sum rules 
in Equation~(\ref{eq:sumrule}) are expressed in terms of magnetic moments 
instead of the expectation values of the spin ($2\left< S_z\right>$) orbital 
($\left< L_z\right>$) or magnetic dipole ($7\left< T_z\right>$) operators that 
are used in some other works (compare Equations~(7) and~(9) of 
Reference~\cite{Sto95}).)

The partial magnetic moments are found by a sum rule analysis as described 
above. The bare results (assuming $P_{\rm circ}=1$ and neglecting $jj$ coupling 
corrections) are $m_s+m_t = 1.73 \: \mu_B$, $m_l = 0.02 \: \mu_B$ for Mn and 
$m_s+m_t = 0.99 \: \mu_B$, $m_l = 0.03 \: \mu_B$ for Co. In the sum rule 
analysis, $n^+_{d, \rm Mn}=4.5$ and $n^+_{d, \rm Co}=2.5$ as found from the 
electronic structure calculations were assumed for the number of $d$ holes in Mn 
and Co, respectively. For both elements, the magnetic dipole or spin anisotropy 
term $m_t$ vanishes in the calculations, as expected for cubic crystals. The 
magnetic moments of Co agree well with the calculations, which predict  
$m_s = 1.00 \: \mu_B$ and $m_l = 0.03 \: \mu_B$.

The sum rule value of the spin magnetic moment of Mn is, however, clearly 
smaller than the predicted value of $m_s = 3.02 \: \mu_B$. It is also evident 
that the total of $\approx(2\times1 + 1.73)\:\mu_B$ found by the sum rule 
analysis does not match the total magnetic moment of about $5\:\mu_B$ found by 
magnetometry. This behaviour is well known from the literature on the $L_{2,3}$ 
XMCD from Mn$^{2+}$. It is usually attributed to $jj$ mixing and overlap of the 
$L_3$ ($2p_{3/2}$) and $L_2$ ($2p_{1/2}$) lines. Goering discusses this in 
detail in Reference~\cite{Goe05}. The spin correction factors were calculated 
according to that work from the present data and are summarised in 
Table~\ref{tab:spincorr}. The branching ratio was calculated either from the 
area ratio $b_i=F_{3}/F_{2}$ or the ratio of the intensity maxima 
$b_m=I^{\max}_{3/2}/I^{\max}_{1/2}$. The {\it mixing} ($X$) and {\it spin 
correction} ($SC$) factors are given by~\cite{Goe05}:

\begin{eqnarray}
     X   & = & \frac{2r-1}{r+1} =   \frac{2-b}{1+b},  \nonumber \\
     SC  & = & \frac{1}{1-2X}   =   \frac{1}{3} \: \frac{1+r}{1-r},
\label{eq:x_sc}
\end{eqnarray}
  
where $r=1/b$ is the inverted branching ratio. The branching ratios are slightly 
higher than the statistical branching ratio of $b_{\rm stat}=2$ as derived from 
the multiplicity of the $p$ states. Therefore, the spin correction factors are 
both slightly less than~1. This means, however, that the spin correction factor, 
in particular that for Mn, does not resolve the discrepancy between the measured 
and calculated magnetic moments. On the contrary, it will decrease the partial 
magnetic moments of both Mn and Co. 

\begin{tablehere}
\tbl{ Spin correction factors for Co and Mn $L_{2,3}$ sum rule analysis. \newline
      The branching ratio $b_m$ calculated from the intensity maxima is 
      given for comparison. The spin correction factors $c_{IE}$ calculated by
      Teramura {\it et al.}~\cite{TTJ96} are also given. 
      }
    {\begin{tabular}{@{}l|ccc|c|cc@{}}
   \toprule
       & \multicolumn{3}{c|}{integral} &         &\multicolumn{2}{c}{Ref.~\cite{TTJ96}} \\
       & $b_i$    & $X$     & $SC$     & $b_m$   & $c_{IE}^{2+}$  & $c_{IE}^{3+}$ \\
   \colrule
    Co & 2.197    & -0.062  & 0.890    & 2.9     & 1.096      & 1.181     \\
    Mn & 2.135    & -0.043  & 0.921    & 3.3     & 1.471      & -         \\
   \botrule
   \end{tabular}}
\label{tab:spincorr}
\end{tablehere}

An empirical correction factor is usually introduced because of difficulties in 
correctly determining the magnetic moment of Mn using sum rule analysis. In many 
publications, the factor $c_{IE}\approx1.5$ based on the work of Teramura {\it 
et al.} is used~\cite{TTJ96}. Their values are also given in 
Table~\ref{tab:spincorr} for Co$^{2+}$, Co$^{3+}$, and Mn$^{2+}$. No value could 
be calculated for Mn$^{3+}$ because $L_3$ and $L_2$ could not be separated. 
Here, the $c_{IE}$ factors of Teramura {\it et al.} result in a spin magnetic 
moment of about 2.6~$\mu_B$ for Mn and thus a total moment of about 4.6~$\mu_B$. 
Together with the uncertainty about the number of core holes, a spin correction 
factor of 1.5 is used in the following for Mn, whereas the Co spin magnetic 
moments are not corrected.

The above reported values concern a single normal incidence measurement. Grazing 
incidence measurements with $\alpha=70^\circ$ were performed to study the 
magnetic anisotropy as proposed by St{\"o}hr and K{\"o}nig~\cite{SKo95}. The 
results are summarised in Table~\ref{tab:angdep}. The normal and grazing 
incidence measurements differ only slightly; thus, no large anisotropies are 
expected.

\begin{tablehere}
\tbl{ Angular dependence of the magnetic moments. \newline
      The spin ($m_s$) and orbital ($m_l$) moments were measured for normal and
      grazing photon incidence and are given in multiples of the Bohr magneton $\mu_B$.
      Mn spin moments are corrected by $c_{IE}=1.5$. }
    {\begin{tabular}{@{}l cc cc@{}}
    \toprule
                     & \multicolumn{2}{c}{Co} & \multicolumn{2}{c}{Mn} \\
                     & $m_s$  & $m_l$         & $m_s$  & $m_l$         \\
    \colrule                                                           
    $m_0^{0^\circ}$  & 0.987  & 0.0345        & 2.492  & 0.0208        \\
    $m_0^{70^\circ}$ & 0.948  & 0.0341        & 2.486  & 0.0202        \\
    \colrule                                                            
    $m_0^\bot$       & 0.987  & 0.0345        & 2.597  & 0.0208        \\
    $m_0^\|$         & 0.942  & 0.0341        & 2.591  & 0.0202        \\ 
    $\Delta m_0$     & 0.044  & 0.0004        & 0.006  & 0.0006        \\
    $m_{av}$         & 0.957  & 0.0342        & 2.593  & 0.0204        \\     
    \botrule
    \end{tabular} }
\label{tab:angdep}
\end{tablehere}

For uniaxial anisotropy, the angular dependence of the 
magnetic moments $m_0^\alpha$ and the angle-averaged moment 
$m_{av}$~\cite{Sto95,Sto99} are given by

\begin{eqnarray}
      m_0^\alpha & = & m_0^\bot \cos^2 \alpha + m_0^\| \sin^2 \alpha,   \nonumber \\
      m_{av}     & = & (m_0^\bot + 2 m_0^\|)/3 \:.
\label{eq:malpha}
\end{eqnarray}

The average moment may be found from a single measurement at the magic angle,
where the second Legendre polynomial becomes zero ($\approx 54.7^\circ$).
The effective moment $m_{\rm eff}$ may be rewritten as the sum of the 
angle-independent spin moment and the angle-dependent dipolar moment 
$m_D$~\cite{Sto95}:

\begin{equation}
	   m^\delta_{\rm eff}=m_s+m_D^\delta \:,
\label{eq:meff}
\end{equation}

where $\delta=\bot$ or $\|$ represent the perpendicular or parallel components 
with respect to the film plane, respectively. The orbital moments are already 
very small; therefore, an analysis of their anisotropy is not reliable. The 
dipolar moments of the Mn atoms are also negligible, as seen from the very small 
difference $\Delta m_0$ (see Table~\ref{tab:angdep}). For the Co atoms, the 
difference $\Delta m_0$ is about 5\%, and the resulting perpendicular and 
parallel components of the dipolar moment are 0.03~$\mu_B$ and -0.015~$\mu_B$, 
respectively. From those values, the out-of-plane component is larger than the 
in-plane component. A nearly vanishing anisotropy is also expected because of 
the small coercive field ($H_c$). From the Stoner--Wolfahrt model, the 
anisotropy energy is on the order of

\begin{equation}
   K = \frac{1}{2}H_A \times \mu_0 M_{\rm sat} \:.
\end{equation}

The present films have $\mu_0H_c=4$~mT ($H_c=0.3$~kA/m), and their magnetic 
moment of $m_s=5$~$\mu_B$ corresponds to a saturation magnetisation of 
$M_{\rm sat}=260$~kA/m. Those values result in an expected anisotropy energy that is 
smaller than 0.5~kJ/m${^3}$, assuming that the coercive field is on the order of 
the anisotropy field ($H_A \approx H_c$). This agrees also with the fact that 
the dipolar term $m_D$ nearly vanishes in accordance with the vanishing of $m_t$ 
in the ab-initio calculations of the electronic structure. 
 
The XMCD spectra were further used to determine the unoccupied electronic 
structure as described in References~\cite{KKS09,KKB09}. The spin-resolved 
unoccupied density of $d$ states is calculated as

\begin{equation}
	n_{d_j}^{\uparrow,\downarrow}(E) = I_0(E) \pm \frac{1}{P_j} I_{\Delta}(E) \:.
\label{eq:pdos}
\end{equation}


$I_0=(I^++I^-)/2$ is the isotropic absorption coefficient (after step-function-
type background subtraction), and $I_{\Delta}=(I^+-I^-)/2$ is half of the 
magnetic dichroism signal. $P_j$ is the spin polarisation obtained for 
excitation from the $p_{3/2}$ ($P_{3/2}=1/4$) and $p_{1/2}$ ($P_{1/2}=-1/2$) 
states (see Reference~\cite{SSi06}, page 391). The upper and lower sign 
correspond to the majority ($n^\uparrow$) and minority ($n^\downarrow$) 
densities, respectively. Note that this is an approximation to the DOS because 
the measured absorption spectra also contain cross section, lifetime, and 
many-particle effects, which are all energy dependent.

Figure~\ref{fig:unoccdos} compares the unoccupied partial $d$ DOS of Co and Mn 
in Co$_2$MnSi. The unoccupied PDOS is derived from the $L_{3,2}$ edges of the 
XMCD data using the spin-resolved unoccupied PDOS function (see 
Equation~(\ref{eq:pdos}) and References~\cite{KKS09,KKB09}). The calculated, raw 
DOS as shown in Figure~\ref{fig:dos} was convoluted by the Fermi-Dirac distribution 
at 300~K and a Gaussian with a width of 300~meV, to account for experimental and 
lifetime broadening. The majority and in particular the minority PDOS of Co and 
Mn are well resolved and in particular the minority PDOS's agree quantitatively 
with the calculated ones. The measured Co PDOS exhibits a shift of the majority 
states with respect to the high density of minority states just above 
$\epsilon_F$. The shift is a consequence of the itinerant $3d-t2_g$ bands that 
dominate the unoccupied majority states at the Fermi energy (see 
Figure~\ref{fig:dos}). The interaction of the core hole in the final state with 
localized $3d$ states lowers the transition energy. The decrease of the 
transition energy is less for itinerant state because they screen the core hole 
to some extent~\cite{Bia82,WCA90}. The different core-hole screening thus 
produces an energy shift between itinerant and localized states. A correlation 
energy of $\Delta E_c=0.5$~eV was suggested in Reference~\cite{KKS09}. An 
increase in the unoccupied majority states at Co is clearly visible at about 
4~eV for both the calculated and measured PDOS. This high density is responsible 
for the {4~eV} satellite observed in the Co $2p$ photo absorption and electron 
emission spectra. The minority PDOS of Mn shows a pronounced maximum at about 
1~eV. In contrast, a characteristic double step increase is observed at the 
measured majority PDOS of Mn. An  energy shift of  $\delta E \approx 1.1$~eV is 
determined from the derivative of the PDOS (see Inset in 
Figure~\ref{fig:unoccdos}). This is in the same order as the shift observed for 
bulk Co$_2$MnSi material and is characteristic of Heusler 
alloys~\cite{KKB09,KBK10,KBA11}. It is induced by electron correlation effects 
between core hole and localised or itinerant minority states. Indeed, the 
correlation energy $\Delta E_c$ may strongly change the minority PDOS determined 
from the XMCD measurements. As a consequence, the half-metallic ferromagnetism 
is not unambiguously detectable with the used method.

\begin{figurehere}
   \centerline{\psfig{file=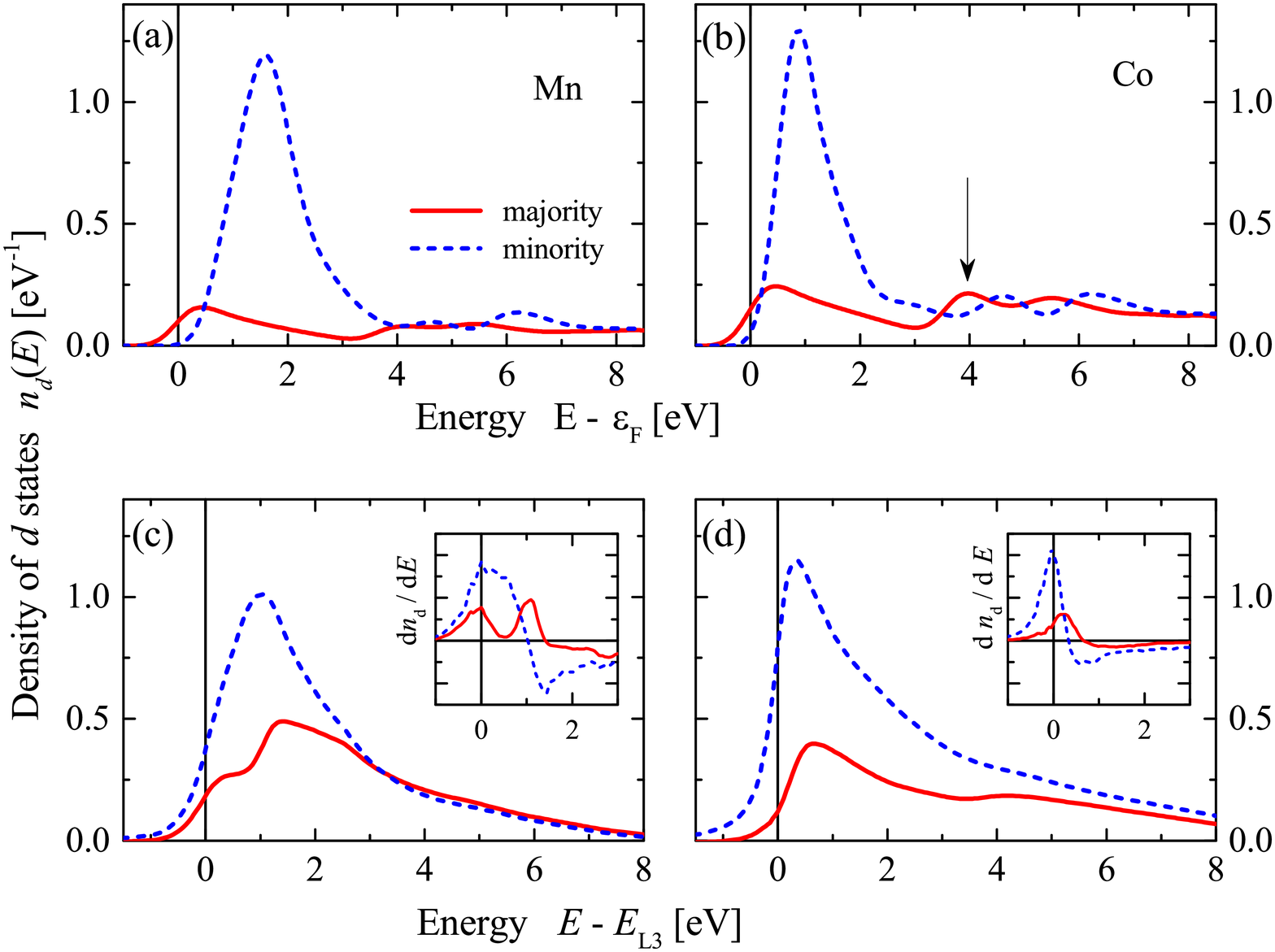,width=8cm}}
   \caption{ (Color online) Unoccupied, spin-resolved partial $d$ DOS of Co and Mn. \\
             Shown are the partial densities of $d$ states (PDOS) from first principles 
             calculations (a, b) and the PDOS taken from the Mn $L_{2,3}$ and Co $L_{2,3}$ 
             absorption data (c, d).
             Insets in (c) and (d) show the derivatives of the experimental data. 
             The calculated PDOS is convoluted by the Fermi-Dirac distribution at 300~K
             and a Gaussian with a width of 300~meV.
             (Note the different zeros ($\epsilon_F$ or $E_{L_3}$) of the energy 
              scales.) }
\label{fig:unoccdos}
\end{figurehere}

\section{Summary and Conclusions} 

The $2p$ core levels of Co and Mn in the HMF Co$_2$MnSi were investigated by 
hard X-ray photoelectron and soft X-ray photoabsorption spectroscopies. Magnetic 
dichroism in HAXPES and XAS were used to explain the states of Co and Mn in thin 
films made of this intermetallic compound.  A combination of {\it ab-initio} 
calculations and core-level spectroscopy with circularly polarised photons 
showed that the Mn $d$ valence electrons and the accompanying magnetic moments 
have a localised character, whereas the Co $d$ valence electrons result in an 
itinerant magnetic moment.

\bigskip
\nonumsection{Acknowledgments}  

\noindent We thank W.~Drube, S.~Francoual, J.~Strempfer, S.~Thiess (PETRA III), 
and the group of R.~Claessen (University of W{\"u}rzburg) for help with the 
setup of the HAXPES experiment at PETRA~III. We are very grateful to the group 
of G.~Reiss (Bielefeld University) for sample preparation. Financial support 
from the DFG-JST (projects P~1.3-A and P~4.8-A in research unit FOR 1464 {\it 
ASPIMATT}) and from the ERC Advanced Grant (291472 Idea Heusler) is gratefully 
acknowledged. HAXPES was performed at beamline P09 of PETRA~III (Hamburg) with 
the support of the Federal Ministry of Education and Research BMBF (Grant Nos. 
05KS7UM1 and 05K10UMA). The XMCD experiment was performed at beamline ID08 of 
the European Synchrotron Radiation Facility (ESRF), Grenoble, France.

\bigskip
\bibliography{cms_dichroism}
\bibliographystyle{unsrt}

\end{multicols}
\end{document}